\documentclass[fleqn,usenatbib]{mnras}

\usepackage{float}
\usepackage{multicol}
\usepackage{graphicx}
\usepackage{subcaption}

\usepackage{ae,aecompl}

\usepackage{graphicx}	
\usepackage{amsmath}	
\usepackage{amssymb}	
\usepackage{url,color}
\usepackage{graphicx}
\usepackage{amsmath,float} 
\usepackage{amssymb}
\newcommand{\ud}{\mathrm{d}}
\newcommand{\pd}{\partial}

\title[Stochastic Core Spin-Up]{Stochastic Core Spin-Up in Massive Stars -- Implications
of 3D Simulations of Oxygen Shell Burning}

\author[McNeill and M\"uller]{
Lucy O.\ McNeill$^{1}$\thanks{E-mail: lucy.mcneill@monash.edu} 
and Bernhard M\"uller$^{1}$
\\
$^{1}$ School of Physics and Astronomy, Monash University, Victoria 3800, Australia
}

\date{Accepted 2020 July 29. Received 2020 July 29; in original form 2020 June 3}

\pubyear{2020}

\begin{document}
\label{firstpage}
\pagerange{\pageref{firstpage}--\pageref{lastpage}}
\maketitle

\begin{abstract}
It has been suggested based on analytic theory that 
even in non-rotating supernova progenitors stochastic
spin-up by internal gravity waves (IGWs) during the late burning stages
can impart enough angular
momentum to the core to result in neutron star birth spin periods below $100 \, \mathrm{ms}$, and a relatively firm upper limit
of $500 \, \mathrm{ms}$ for the spin period. We here investigate this process using a 3D simulation of oxygen shell burning in a $3M_\odot$ He star.
Our model indicates that stochastic spin-up by IGWs is
less efficient than previously thought. We find that the stochastic 
angular momentum flux carried by waves excited at the shell boundary
is significantly smaller for a given convective luminosity 
and turnover time than would be expected from simple dimensional 
analysis. This can be explained by noting that the waves launched by
overshooting convective plumes contain modes of opposite angular
wave number with similar amplitudes, so that the net angular
momentum of excited wave packets almost cancels. We find that
the wave-mediated angular momentum flux from the oxygen shell
follows a random walk, but again dimensional analysis
overestimates the random walk amplitudes since the correlation
time is only a fraction of the convective turnover time. Extrapolating
our findings over the entire life time of the last burning stages
prior to collapse, we predict that the core angular momentum from 
stochastic spin-up would translate into long birth spin periods of 
several seconds for low-mass progenitors and no less than $100\, 
\mathrm{ms}$ even for high-mass progenitors.
\end{abstract}



\section{Introduction}
The birth spin periods of pulsars potentially provide a window into
the inner workings of angular momentum transport in massive stars
and of the core-collapse supernova explosion mechanism. The bulk
of the observed solitary neutron star population has birth spin periods of hundreds of ms, and some as fast as tens of ms \citep{Fau2006, Popov2010, Popov2012, Gullon2014}. Slow birth spin periods of several hundred ms can be inferred from age and spin down measurements (e.g., PSR J1955+5059 in
\citealp{Noutsos2013}), but without independent measures of the age 
and braking index, the slow end of the population cannot be tightly 
constrained. Even for young pulsars
with spin periods of $\mathord{>}100\, \mathrm{ms}$ (e.g. PSR 
J0248+6021, which has a spin period of $217\,\mathrm{ms}$ and age 
$62\, \mathrm{kyr}$ \citealp{Theureau2011}), uncertainties in
possible instabilities and spin-down mechanisms during the first
years means that one needs to be careful in
drawing conclusions from such low spin rates
on the birth spin period. Despite 
long-standing efforts to understand the pre-collapse spin periods
of massive stars \citep[e.g.,][]{Heger2000,hirschi_04,Heger2005} and 
spin-up/spin-down processes during the supernova
explosion \citep{Blondin2007,Rantsiou2011,Wongwathanarat2013,Kazeroni2016,BM2019a,stockinger_20}, it is still
not clear what shapes the observed period distribution.

In particular, there are still considerable uncertainties concerning
the angular momentum transport processes in stellar interiors that
determine the pre-collapse rotation profiles of massive stars.
Stellar evolution models incorporating magnetic torques \citep[e.g.,][]{Heger2005,Suijs2008,Cantiello2014} 
from the Tayler-Spruit dynamo \citep{Spruit2002} have long defined the
state of the art for the treatment of stellar rotation in massive stars.
In recent years, however, asteroseismology has revealed
unexpectedly slow core rotation rates in low-mass red giants
\citep{Beck2012, Beck2014, Deh2012, Deh2014,Mosser2012},
which cannot be explained by the classic Tayler-Spruit dynamo,
suggesting that even more efficient angular momentum transport
mechanisms must operate in nature \citep{Fuller2014,Cantiello2014,Wheeler2015}.

Different solutions have been proposed to resolve this problem.
\citet{Fuller2019} have proposed a modification of the Tayler-Spruit dynamo
which efficiently slows down progenitors to be roughly consistent with both the
observed neutron star birth spin periods and, unlike the Tayler-Spruit dynamo, the core rotation rates of red
 giants. It cannot, however, explain the rotation rates of intermediate mass stars \citep{Denhartough2020}. As an alternative solution that does not rely on magnetic torques,
many studies have investigated angular momentum transport by internal gravity
waves (IGWs) in low- and high-mass stars during various evolutionary phases
\citep[e.g.,][]{Zahn1997,Kumar1999,Talon2003,Charbonnel2005,Rogers2006,Fuller2014,Belkacem2015,Pincon2016,Pincon2017}
and suggested that IGWs can maintain strong core-envelope coupling only up to
the subgiant branch. 

Angular momentum transport by IGWs could, however, again play an important
role during more advanced burning stages in massive stars because the enormous
convective luminosities and relatively high Mach numbers lead to a strong
excitation of IGWs at convective boundaries \citep{Fuller2015}. The familiar
wave filtering mechanism for prograde/retrograde modes could then effectively
limit the degree of differential rotation in the interior and spin down the
core on rather short time scales if rotation is fast to begin with
\citep{Fuller2015}. The authors find that for initially rapidly rotating
progenitors, angular momentum transport via g-modes is not efficient enough to explain the
pulsar population. The authors find a maximum spin frequency bound from spin--down of $\sim 3$ms, which is consistent with the range of possible birth spins for the fastest spinning solitary pulsar (\citealp{Marshall1998}, spin period of 16 milliseconds today).
Additionally, \citet{Fuller2015}, pointed out that IGWs could lead to a
stochastic core \textit{spin-up} in initially non-spinning progenitors. The
idea here is that stochastically excited IGWs from the convective boundaries
of the Si, O, or C shell propagate into the core, where they break and deposit
their randomly varying angular momentum. The core angular momentum
executes a random walk, as long as there is an influx of IGWs from strong
burning shells. In principle,
IGWs can also be excited by core convection or shells that end up in the core,
and carry angular momentum \emph{outwards}. This mirror process was found
to be subdominant in the case studied by \citet{Fuller2015}, however. 

Rotation could potentially modify this random-walk process in a non-trivial
way. Once a substantial rotation rate is present in stars, the wave-filtering process \citep{Talon2005} can change the direction of zonal mean flows, similar to the ``quasi-biennnial'' oscillations observed in the Earth's atmosphere, where winds at the equator completely change direction (e.g., prograde to retrograde) every 14 months \citep{Baldwin2001}. Changes driven by IGWs have been studied in the context of the Sun with varying success \citep{Kumar1999, Kim2001,Talon2002, Rogers2006}. These studies model the evolution of solar zonal velocity fields over year-to-decade timescales, motivated by the observed 11-year sunspot cycle and the solar wind's 1.3 year oscillation cycle. Unlike the Sun, during late burning phases the shell structure changes on timescales of minutes, and it is unclear whether we can even expect similar long-time oscillations on secular
time scales to exist. If they do, identifying them will be challenging in 3D simulations due to difficulties in identifying eigenmodes for a non--stationary background.

\citet{Fuller2015} argue that the stochastic spin-up will result
in typical neutron birth periods scattering over a range
between about $40\, \mathrm{ms}$
up to an effective upper limit
of $\mathord{\sim} 500\,\mathrm{ms}$
without the need to assume any
progenitor rotation. If the core angular momentum is
indeed determined by this stochastic spin-up process,
this could explain the paucity of long spin periods
of $\mathord{\gtrsim} 500\,\mathrm{ms}$ among young neutron stars.
Aside from the implications for neutron star spin periods, the excitation
of IGWs during the late convective burning stages is also of interest
because it could drive mass loss in supernova progenitors shortly before
collapse \citep{QS2012,Fuller2017} or lead to envelope inflation
\citep{Mcley2014}. This could explain observations of circumstellar
material (CSM) from late pre-collapse outbursts in a significant number of
observed supernovae \citep[for an overview, see][]{Foley2011,Smith2014,Bilinski2015,Smith2017}, although other mechanisms such as 
flashes from degenerate shell burning \citep{Woosley2015}
or the pulsational pair instability \citep{Woosley2007} may be required to 
account for more spectacular cases with several solar masses of CSM in
Type~IIn supernovae.

In this paper, we further investigate the excitation of IGWs during late-stage
convective burning and its implication for stochastic core spin-up. Different
from the problem of wave excitation by convection during early evolutionary
stages, this problem cannot be addressed by asteroseismic measurements. The
extant studies of \citet{QS2012,Fuller2015,Fuller2017} 
strongly rely
on the analytic theory of wave excitation by turbulent convection that has
been developed over decades \citep{Lighthill1952,Townsend1966,GK1990,LQ2013}. 
Numerical
simulations of IGW excitation at convective boundaries have been conducted
for earlier evolutionary stages 
in 2D \citep{Rogers2013}
and 3D \citep{Alvan2014,Alvan2015,Edelmann2019}, 
but cannot be easily extrapolated to the late,
neutrino-cooled burning stages where the P\'eclet number is lower and
compressibility effects play a bigger role.
During the early evolutionary stages, asteroseismology
can be used to directly test the underlying theories for IGW excitation
by convection \citep{Aerts2015,Aerts2017b}.
 On the other hand, a number of
studies have already addressed the late burning stages in 3D
\citep[e.g.,][]{Meakin2007,Arnett2008,BM2016,Jones2017,Cristini2017,Andrassy2018,BM2019a,Yadav2019},
but without addressing the problem of stochastic spin-up.

In this study, we use a full $4 \pi$--3D numerical simulation to quantitatively
study the idea of stochastic spin-up for the first time. We consider a $3
M_\odot$ He star model from \citet{BM2019a}, which we evolved for
7.8 minutes, or $\mathord{\sim} 35$ convective turnover times. We
consider the excitation of waves by vigorous convection in the oxygen burning
shell of this model to check the assumptions underlying the theory of
\citet[F15 henceforth]{Fuller2015} for stochastic core spin-up.
We focus on the outward flux of angular momentum 
from the oxygen shell since
wave excitation at the outer boundaries of a convective
shell or core is more tractable from the numerical point
of view than the excitation of inward-propagating
waves from the inner boundary of a convective shell.
Since the physical process of wave excitation is the
same, this approach nonetheless allows us to study
the process in the relevant physical regime and
check the scaling relations and dimensionless parameters
used in the theory of F15.

The paper is structured as follows: In Section~\ref{sec:theory}, we review key
elements of the theory of F15 for the stochastic spin-up of supernova
progenitor cores by IGWs. In Section~\ref{sec:setup} we describe the 3D
progenitor model. 
In Section~\ref{sec:results} we
analyze the simulation using spherical Favre decomposition to determine the
energy and angular momentum flux in convectively stable zones. We compare
these results with the random walk model of \citet{Fuller2015}, and finally
comment on the implications of our results for neutron star birth spins. In Section~\ref{sec:ang_cons}
we discuss potential numerical issues that may affect
our results, such as numerical angular momentum
conservation errors, and then 
summarise our findings in Section~\ref{sec:conclusions}.

\section{Theory of Stochastic Spin-Up}
\label{sec:theory}
Gravity waves ($g$-modes) are generated by turbulent convection at convective
boundaries, and propagate in the neighboring convectively stable
regions. The theory of F15 relies on a few key assumptions about
the wave excitation process to model the stochastic spin-up of the core in
non-rotating or slowly rotating progenitors.

The first of these assumptions concerns the (time-averaged) IGW energy
flux $\dot{E}$ in the stable regions. Following established analytic
theory for $g$-mode excitation 
\citep{LQ2013},
F15 assume that the wave energy flux $\dot{E}$ 
depends on the convective luminosity $L_\mathrm{conv}$ in the driving motions
and the convective Mach number $\mathcal{M}$,
\begin{equation}
\label{eq:e_wave}
\dot{E}=
L_\mathrm{conv} \mathcal{M}^\alpha,
\end{equation}
where the power-law exponent $\alpha$ is expected to lie
in the range $\alpha=5/8\texttt{-}1$.

While the wave energy flux can be assumed to vary only mildly in time, the
overshooting convective plumes will continuously excite different modes, and
the overall angular momentum flux $\dot{\mathbf{J}}$ carried by the waves will
fluctuate according to the wave numbers of the excited modes. F15 assume that
the direction of the angular momentum flux vector remains correlated over
roughly one convective turnover time $\tau$ and varies randomly on longer time
scales, so that time-integrated spin-up
 $\Delta J =\int\dot{\mathbf{J}}\,\ud t$ can be approximated as the result of a random walk by
the typical angular momentum $\delta J$ of a correlated wave packet
per time $\tau$. Thus, after $\mathcal{N}$ convective turnover times,
the expectation value $\langle \Delta J^2 \rangle$ is
\footnote{Properly speaking, each component $\Delta J_i$ will execute 
an independent random by with step size
$\delta J_i=\delta J/ \sqrt{3}$ if the angular momentum of the
wave packets is randomly oriented, but the factor $1/\sqrt{3}$
cancels again when we consider the expectation value
of $\Delta J^2$ since
$\langle \Delta J^2\rangle =3 \langle \Delta J^2\rangle$.
}
\begin{equation}
 \langle \Delta J^2 \rangle
= \mathcal{N} \delta J^2.
\end{equation}

At this stage, one still needs to specify the typical angular momentum $\delta
J$ carried by one correlated wave packet. For a single mode of spherical
harmonics degree $\ell$ and order $m$, the $z$-component $\dot{J}_z$ of the
flux of angular momentum is related to the wave energy flux as
\citep{Zahn1997,Kumar1999},
\begin{equation}
\dot{J}_z=\frac{m}{\omega}\dot{E},
\label{eq:GKE}
\end{equation}
where $\omega$ is the mode frequency. F15 suggest replacing $m$ and $\omega$
by appropriate averages $\bar{m}$ and $\bar{\omega}$ for the wave packet so that
\begin{equation}
\dot{J}_\mathrm{F15}=\frac{\bar{m}}{\bar{\omega}}\dot{E}
=\frac{\bar{m}}{\bar{\omega}}\mathcal{M}^\alpha\, L_\mathrm{conv}.
   \label{eq:RWJ}
\end{equation}
The authors argue that the excited modes will predominantly have periods of the order
of the convective turnover time so that $\bar{\omega}=2\pi/\tau$.
For the average wave number, F15 choose $\bar{m}=1$,
arguing that this is the most conservative estimate provided that each wave packet consists
only of modes of positive or negative $m$. We shall revisit this
assumption later in Section~\ref{sec:results}.

Integrating over one turnover time and using 
Equation~(\ref{eq:e_wave}), one then finds
\begin{equation}
  \delta J_\mathrm{F15} = \frac{\bar{m} \tau^2 \dot{J}_\mathrm{F15}}{2\pi}
= \frac{\tau^2 \mathcal{M}^\alpha L_\mathrm{conv}}{2\pi},
\end{equation}
and the expected spin-up after $\mathcal{N}$ turnovers is
\begin{equation}
	\langle \Delta J_\mathrm{F15}^2 \rangle^{1/2} = 
\sqrt{\mathcal{N}}\delta J_\mathrm{F15}
=
 \frac{\sqrt{\mathcal{N}} \bar{m} \tau^2
	\dot{J}_\mathrm{F15}}{2\pi\sqrt{3}} = \frac{\sqrt{\mathcal{N}} \bar{m} \tau^2
	\mathcal{M}^\alpha L_\mathrm{conv}}{2\pi\sqrt{3} } 
	\label{eq:RWMa},
\end{equation}
where the factor $1/\sqrt{3}$ accounts for the random orientation
of the vectorial angular momentum carried by different wave packets.

\section{Shell Burning Simulation}
\label{sec:setup}
\begin{figure}
\centering
\begin{subfigure}[b]{\linewidth}
  \includegraphics[width=\linewidth]{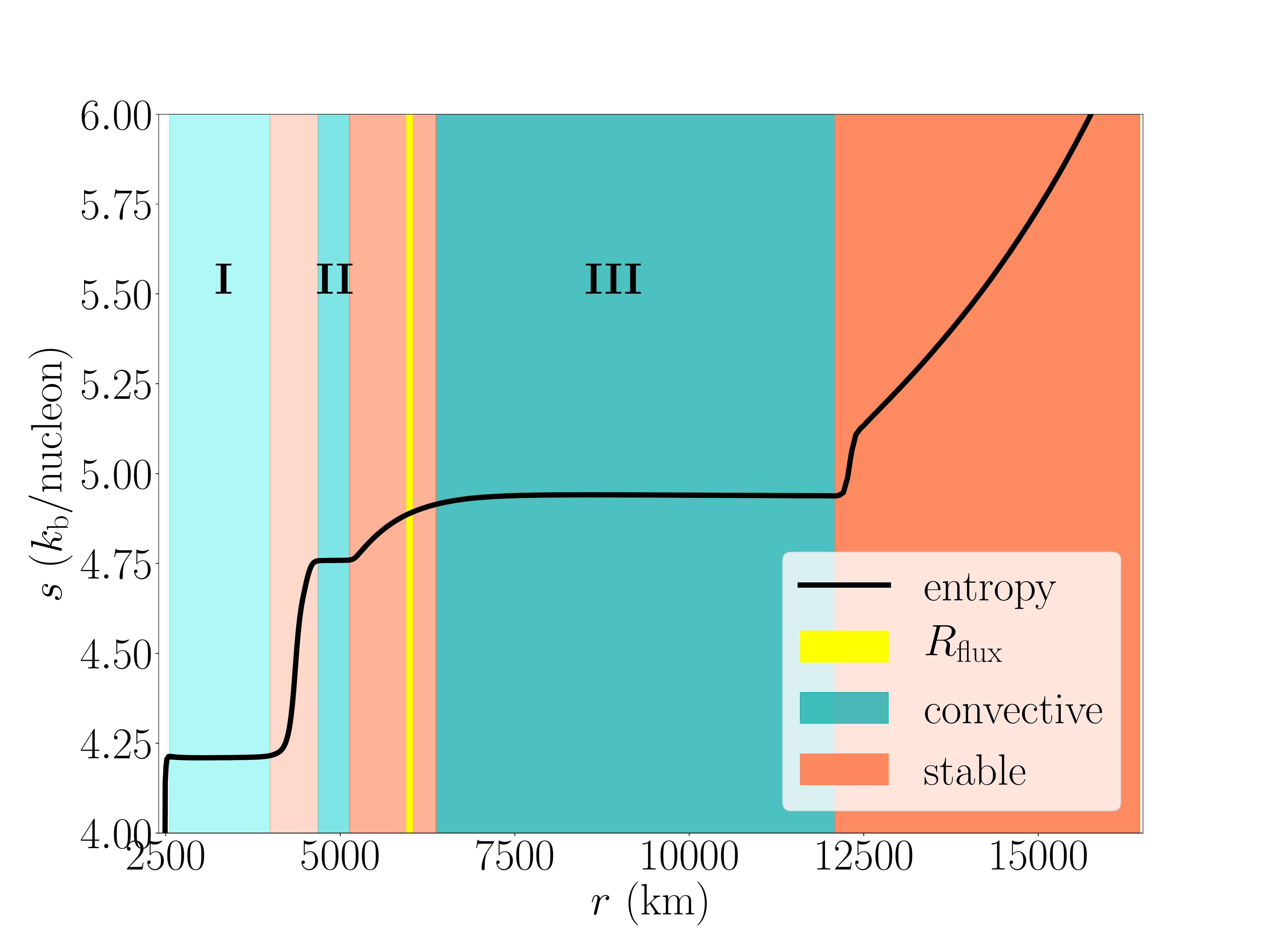}
\caption{Shell structure at early times.}
\label{fig:midpoints} 
\end{subfigure}
\begin{subfigure}[b]{\linewidth}
\includegraphics[width=\linewidth]{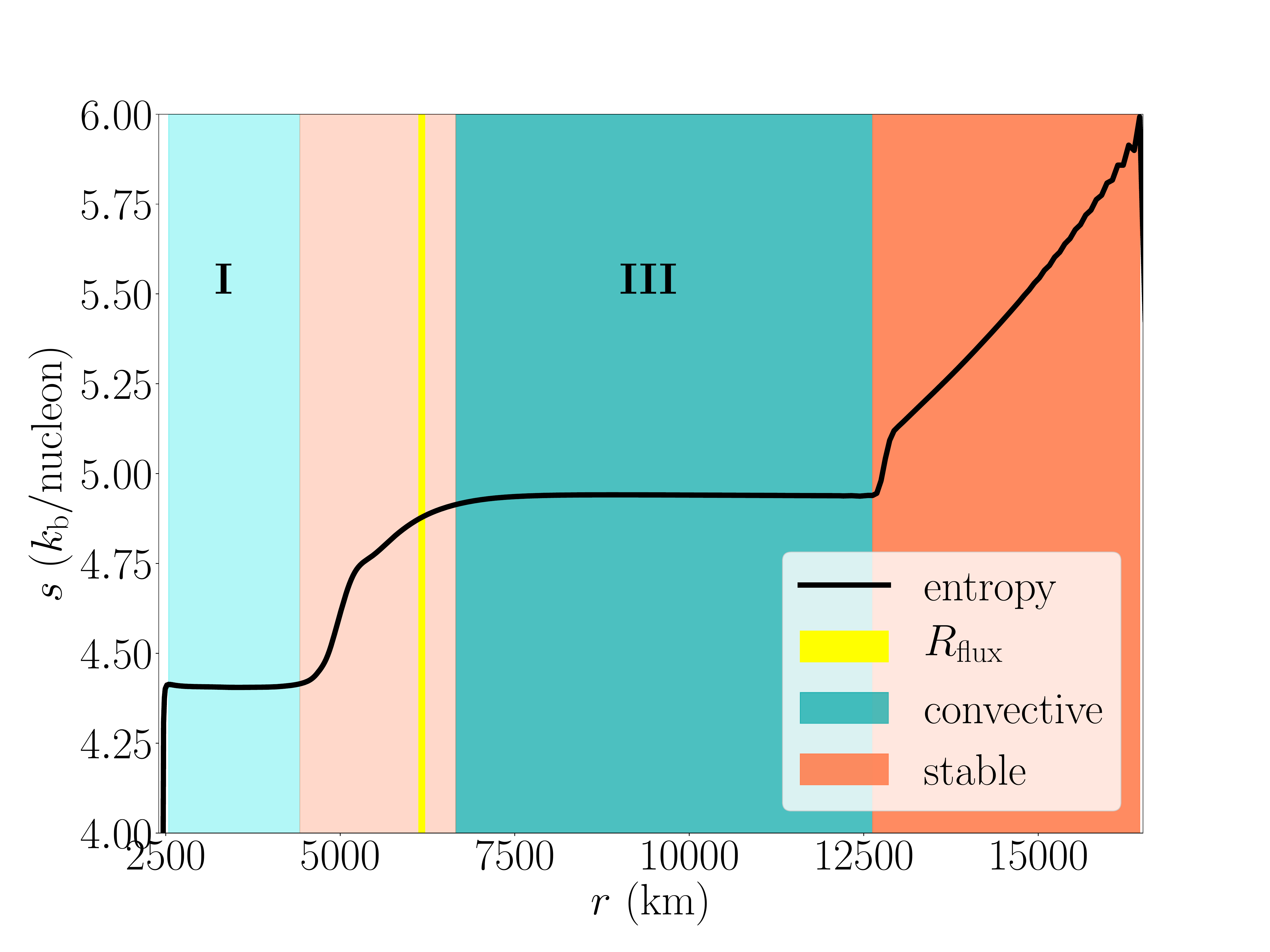}
\caption{Shell structure at late times, after the thin neon burning region (II) has disappeared.}
\label{fig:lateS}
\end{subfigure}
\caption{Spherically averaged 
profiles of the specific entropy $s$ (solid black), against our defined regions at early ($t=144\, \mathrm{s}$) and late ($t=368\, \mathrm{s}$) times. Convective (turquoise) and convectively stable (coral) regions are labelled (I)--(III), with increasing darkness from the inner boundary to the outer boundary.
Turbulent fluxes between the convective regions~(I) and (III) are evaluated at the radius $R_\mathrm{flux} =R_{m=1.65 M_\odot}$, which is in the convectively stable region between zones (II) and (III) up until around $400\, \mathrm{s}$. The second convective region disappears completely at $344\, \mathrm{s}$; the shell structure after
the disappearance of region~(II) is shown in Figure~\ref{fig:lateS}.
}
\label{fig:earlyS}
\end{figure}

The challenge is now to determine whether
the key approximations in Equations~(\ref{eq:e_wave}), (\ref{eq:RWJ}), and
(\ref{eq:RWMa}) are borne out by full 3D simulations.
We analyse a 3D model of convective shell burning in a
non-rotating $3 M_\odot$ helium star from \citet{BM2019a}. Following
the methodology of \citet{BM2016}, we map the the underlying 1D stellar
evolution model from the \textsc{Kepler} code
\citep{weaver_78,heger_10} into the finite-volume code \textsc{Prometheus}
\citep{fryxell_89,fryxell_91}
at a late pre-collapse stage
about 8~minutes before the onset of collapse.
Our simulation includes the mass shells initially
located between $2,420\,\mathrm{km}$ and $16,500\, \mathrm{km}$, which covers
three distinct convective zones as illustrated (in turquoise) by the progenitor entropy
profile in Figure~\ref{fig:midpoints}.
Initially there are three convective burning zones: (I), (II) and (III), which are oxygen, neon, and carbon shells respectively.
The neon burning region~(II) is consumed
before collapse and is so thin that $g$-modes
will in fact be able to propagate through it 
for much of the simulation.
By $344\, \mathrm{s}$ (5.77 minutes) region~(II) has 
disappeared completely, and there is just
one convectively stable region between (I) and (III), shown in Figure~\ref{fig:lateS}.
The inner boundary of the grid is contracted
in accordance with the mass shell trajectory from the 1D \textsc{Kepler}
model.
We use uniform spacing in $\log r$ for the radial coordinate
and the axis-free overset
Yin-Yang grid \citep{kageyama_04} as implemented
by \citet{melson_15a} to cover the full sphere. The grid
resolution is $400 \times 56 \times 148$ zones in radius $r$, latitude
$\theta$, and longitude $\varphi$
on each of the Yin-Yang patches, which corresponds to an angular
resolution of $2^\circ$. Extremum-preserving 6th-order
PPM reconstruction has been implemented in \textsc{Prometheus}
following \citet{colella_08}.

\label{section:progenitor}

\section{Results}
\label{sec:results}

\subsection{Visual identification of excited modes}
\label{sec:flow}
We run the model for 7.8 minutes during oxygen burning. It takes about one minute for convection to
fully develop in region~(I) from the 1D initial conditions, so we only consider the phase
\textit{after} the first minute in our subsequent analysis.

\begin{figure*}
\begin{multicols}{2}
\includegraphics[width=\linewidth]{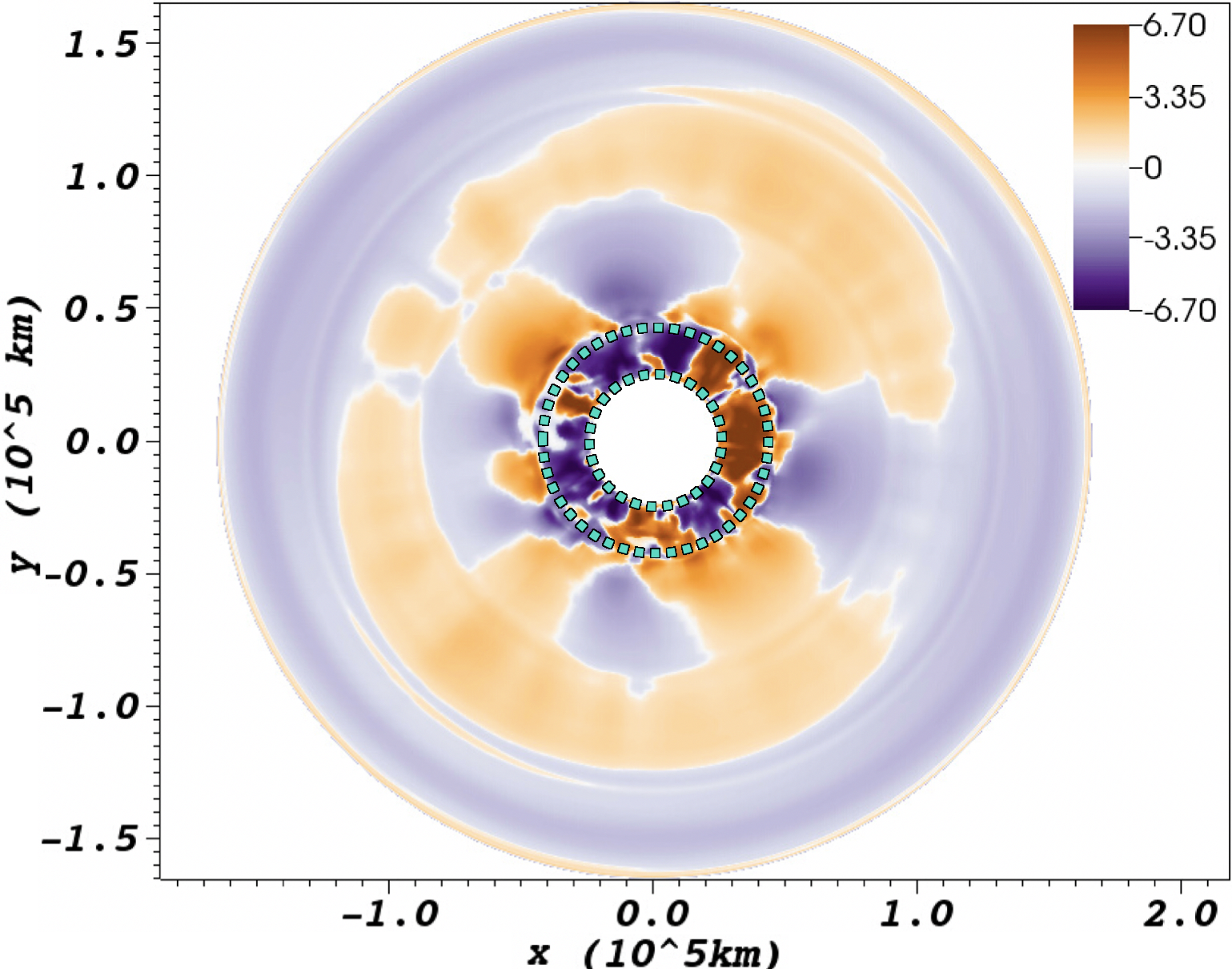}\par 
  \includegraphics[width=\linewidth]{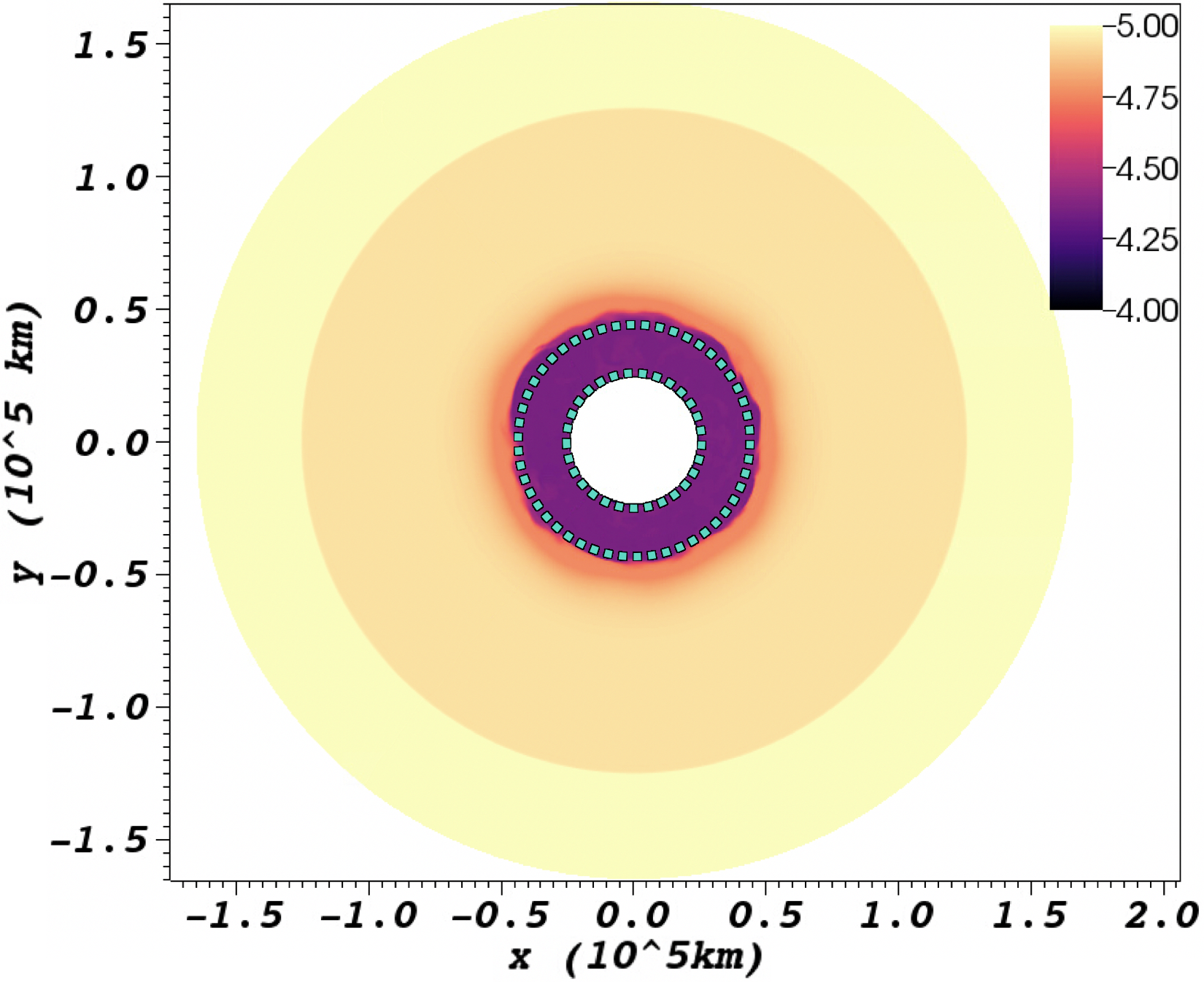}\par 
  \end{multicols}
\caption{2D slices showing the cube root $v_r^{1/3}$ of the radial velocity $v_r$ in $\mathrm{km}\, \mathrm{s}^{-1}$
(left) and the entropy $s$ in $k_\mathrm{b}/\mathrm{nucleon}$
(right) at a time of $327\, \mathrm{s}$, where $v_r^{1/3} = 3.35
\, \mathrm{km}^{1/3}\, \mathrm{s}^{-1/3}$ and $6.70\, \mathrm{km}^{1/3}\, \mathrm{s}^{-1/3}$ correspond to $v_r = 40\, \mathrm{km}\, \mathrm{s}^{-1}$ and $300 \mathrm{km}\, \mathrm{s}^{-1}$ respectively}. In region~(I) (demarcated by dotted turquoise lines)
we find turbulent motions of high velocity. In the surrounding
stable region outside $r\approx 4400\, \mathrm{km}$, we find
coherent flow patterns dominated by moderately high multipoles,
which is indicative of $g$-mode activity.
\label{fig:visit}
\end{figure*}

In Figure~\ref{fig:visit} we show 2D slices of the cube root\footnote{We plot the cube root to better identify flow features with small velocities.} of the radial 
velocity $v_r^{1/3}$ and specific entropy $s$. We have restricted the 
 velocity scale 
corresponding to $\texttt{-}300 \, \mathrm{km}\, \mathrm{s}^{-1} < v_r < 300\, \mathrm{km}\, \mathrm{s}^{-1}$ ($\texttt{-}6.75 \, \mathrm{km}^{1/3}\, \mathrm{s}^{-1/3}< v_r^{1/3} < 6.75 \, \mathrm{km}^{1/3}\, \mathrm{s}^{-1/3}$)
to show both the convective motions in region~(I) and the
excited waves in the surrounding stable region.
The boundary between region~(I) enclosed by turquoise dotted lines and the stable region
is clearly identifiable in the entropy plot as a
relatively sharp discontinuity at a radius
of about $4400\, \mathrm{km}$; it is evident
that the convective plumes do not overshoot strongly into
the stable regions. This clearly identifies the 
slower, laminar motions of a few $10\, \mathrm{km}\, \mathrm{s}^{-1}$ ($v_r^{1/3}=3.35 \, \mathrm{km}^{1/3}\, \mathrm{s}^{-1/3}$)
as excited modes. We note that these wave patterns do not appear spherical, as would be expected for waves in lower Mach number flows \citep[e.g.,][]{Horst2020}), where the timescale for convective forcing is comparable to the propagation speed. Here, $\mathcal{M}\sim 0.1$ and the convective velocities reach a few $100\, \mathrm{km}\, \mathrm{s}^{-1}$ ($v_r^{1/3}=6.70$). The excited modes are therefore of moderately
high wave number and clearly not dominated by dipole
or quadrupole modes.

\subsection{Expected IGW frequencies}
 We compute the Brunt--V\"ais\"al\"a frequency $\omega_\mathrm{BV}$ in the convectively stable regions using
\begin{equation}
  \omega_\mathrm{BV}^2 = g \left( \frac{1}{\Gamma}\frac{\mathrm{d \ ln} P}{\mathrm{d}r} - \frac{\mathrm{d \ ln}\rho}{\mathrm{d}r}\right),
  \label{eq:BV}
\end{equation}
where $g$ is the local gravitational acceleration
\begin{equation}
  g(r) = \frac{G M(r)}{r^2},
\end{equation}
and $\Gamma$ is the adiabatic exponent $\Gamma = (\partial \mathrm{ln} P/ \partial \mathrm{ln} \rho)_s = \rho/P c_s^{2}$ in terms of the sound speed $c_s$, the pressure $P$, and the density $\rho$ at radius $r$.
\begin{figure}
 \centering
 \includegraphics[width=0.5\textwidth]{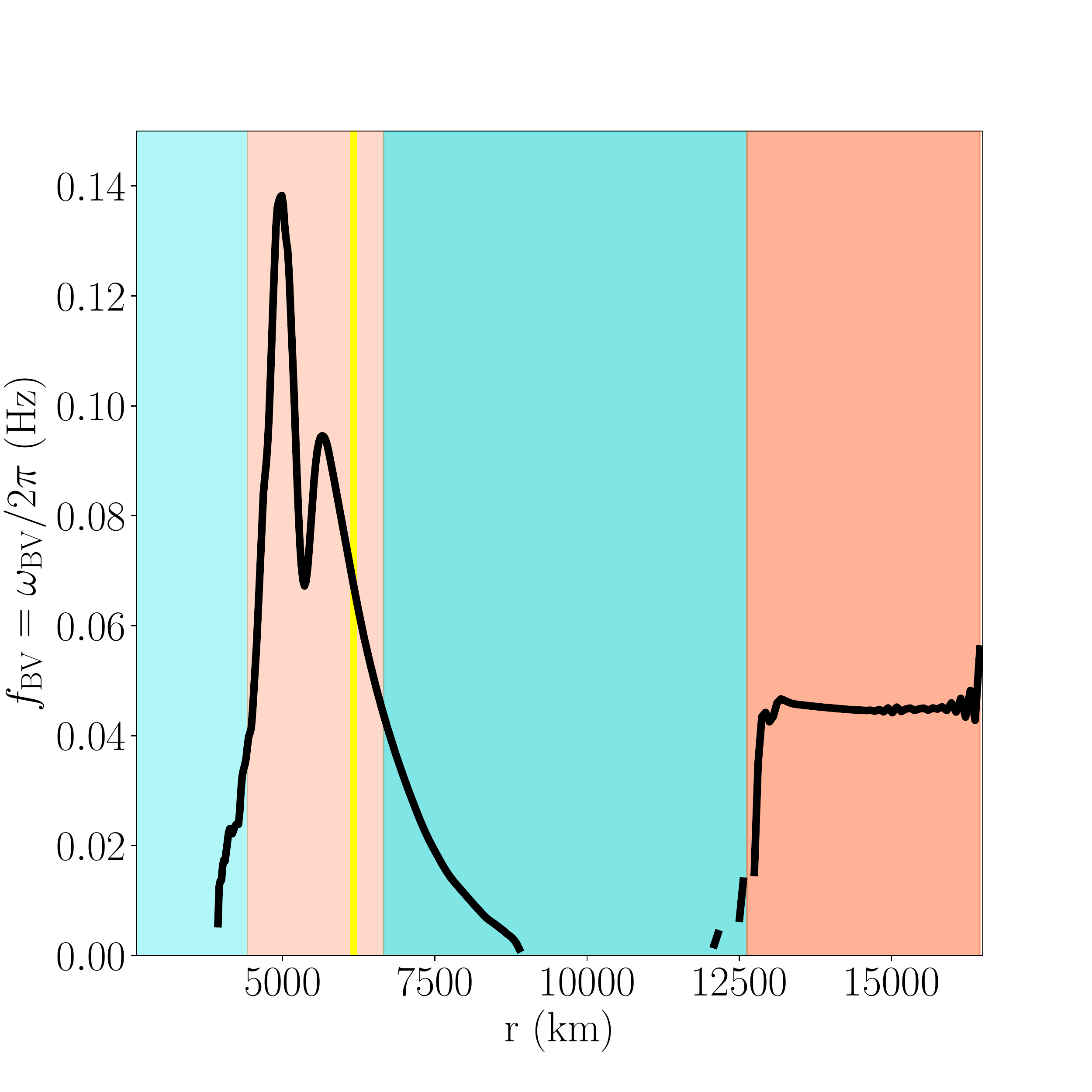}
 \caption{Brunt--V\"ais\"al\"a frequency $\omega_\mathrm{BV}$
 as a function of radius $r$
 at $368\, \mathrm{s}$ (same as Figure \ref{fig:lateS}), after the middle convective region has disappeared. Based on this profile, gravity waves may propagate in the coral convectively stable zones 
 at frequencies below $\mathord{\sim} 0.1\, \mathrm{Hz}$.
 }
 \label{fig:BV}
\end{figure} 
In Figure \ref{fig:BV} we plot $f_\mathrm{BV} = \omega_{\mathrm{BV}}/2\pi$ for the simulated region as a function radius at a late time of $368\, \mathrm{s}$ (same time as Figure \ref{fig:lateS}). In the first convectively stable region, 
$f_\mathrm{BV}$ peaks between $0.1\texttt{-}0.4\, \mathrm{Hz}$ during the whole simulation. After about 
$350\, \mathrm{s}$ it stays at $\mathord{\sim} 0.1\,\mathrm{Hz}$. Waves with frequencies lower than these may propagate as gravity waves in this region. Convective and convectively stable regions (defined by the entropy gradient of Figure \ref{fig:lateS}) at this time step are included with the same colour scheme as Figure~\ref{fig:earlyS}. The expected 
mode frequencies are of a similar order as the convective
turnover frequency, and hence the conditions for
IGW excitation by convective motions should be close to optimal.

Note that, strictly speaking, the Brunt--V\"ais\"al\"a frequencies computed with Equation~(\ref{eq:BV}) are only valid in the linear approximation, where the runaway (convectively unstable) or oscillation (convectively stable) of a stochastically displaced fluid parcel is determined by the local buoyancy excess upon displacement. The fact that there are real Brunt--V\"ais\"al\"a frequencies outside of our convectively stable regions (defined by the entropy gradient) can be explained by the deceleration of fluid parcels as they approach the boundary, where they can penetrate (overshoot) it (see \citealp{Mocak2008} for a detailed explanation).

\subsection{Angular momentum flux in excited waves}

\label{section:reynolds}
In order to quantify the energy and angular momentum flux carried by the waves
excited a convective boundaries, we make use of a spherical Reynolds/Favre
decomposition of the flow \citep{Favre1965}. We use hats (or, alternatively,
angled brackets) and primes for the volume-weighted Reynolds averages and
fluctuating components of extensive quantities such as the density $\rho$ and
pressure $P$ (e.g., $\hat{\rho}$ and $\rho'$). These are defined for any such
quantity $X$ as
\begin{eqnarray}
 \hat{X}(r) &=& \langle  X \rangle =  \int X \, d \Omega,
\\
 X'(r,\theta,\varphi)&=&X-\hat{X}(r).
\end{eqnarray}
Mass-weighted Favre averages and fluctuating components of
intensive quantities like the internal energy density $\epsilon$
are denoted by tildes and double primes (e.g., $\tilde{\epsilon}$, and
$\epsilon''$). For any such quantity $Y$, we have
\begin{eqnarray}
 \tilde{Y} (r) &=& \frac{ \int \rho Y 
\, d \Omega}{\int \rho
\, d \Omega},
\\
Y''(r,\theta,\varphi)&=&Y-\tilde{Y}(r).
\end{eqnarray}

Disregarding subdominant terms for the work done by turbulent Reynolds
stresses, the Favre-averaged energy equation reads
\citep{BM2019b},
\begin{equation}
\label{eq:favre}
\begin{split}
& \frac{\partial}{\partial t} \left( \hat{\rho} \tilde{\epsilon} 
+ \hat{\rho} \frac{|\tilde{\mathbf{v}}|^2}{2}
+ \hat{\rho} \frac{\langle |\tilde{\mathbf{v}}''| \rangle^2}{2}\right) 
+ \nabla \cdot \left[ \left( \hat{\rho} \tilde{\epsilon} 
+ {\hat{\rho}\frac{|\tilde{\mathbf{v}}|^2}{2}}
+\hat{P} \right)\tilde{\mathbf{v}} \right] \\
 & 
 + \nabla \cdot \left[ 
\langle{\hat{\rho} \epsilon'' \mathbf{v}''
    \rangle }
 + \langle P' \mathbf{v}'' \rangle 
 + \langle \rho \mathbf{v}'' \frac{|\mathbf{v}''^2|}{2}\rangle \right]
=0. 
\end{split}
\end{equation}
The relevant terms containing the energy flux carried by waves excited at
convective boundaries are the convective energy flux $\mathbf{F}_\mathrm{conv}
= \langle \hat{\rho} \epsilon'' \mathbf{v}''\rangle$ (for the turbulent
transport of internal energy), the acoustic energy flux
$\mathbf{F}_\mathrm{sound} = \langle P' \mathbf{v}'' \rangle$, and the
turbulent kinetic energy flux $\mathbf{F}_\mathrm{kin} = \langle \rho
\mathbf{v}'' \frac{|\mathbf{v}''^2|}{2}\rangle$. 

Strictly speaking, however, one can only properly
separate the flux carried by gravity waves
and acoustic waves using a decomposition of 
the fluctuating components into eigenmodes.
When using the Favre decomposition
of the energy equation, the wave energy
flux is split between the three
turbulent flux components. At
the same time acoustic waves and 
entrainment contribute to the turbulent
fluxes, which is particularly problematic
in the case of entrainment, which
produces a \emph{negative} energy
flux near the convective boundary.

Due to these difficulties in isolating the wave energy flux $\dot{E}$ from region~(I), we do not test Equation~(\ref{eq:e_wave}), for
which there is strong justification
from theory and simulations
anyway \citep{GK1990,LQ2013,Pincon2016}.
Instead, we assume $\dot{E} = \mathcal{M}
L_\mathrm{conv}$, and directly
test the dependence of
$\dot{J}$ on $L_\mathrm{conv}$ and
$\mathcal{M}$ in Equation~(\ref{eq:RWJ}).

To obtain $\dot{J}$, we consider the
Favre-averaged equation for the transport of angular momentum, which
only contains two flux terms 
$\mathbf{F}_{\mathrm{adv}}$ and
$\mathbf{F}_{\mathrm{turb}}$ for the
mean (advective) and turbulent angular momentum flux,
\begin{equation}
\frac{\pd \langle \rho \mathbf{l} \rangle}{\pd t}
+
\nabla\cdot \mathbf{F}_\mathrm{adv}
+
\nabla\cdot \mathbf{F}_\mathrm{turb}
=0,
\end{equation}
where
\begin{equation}
\mathbf{F}_{\mathrm{turb}} = \langle \rho \mathbf{l}'' u_r''\rangle,
\end{equation}
and
\begin{equation}
\mathbf{F}_{\mathrm{adv}} = 
\langle \rho \tilde{\mathbf{l}} \tilde{u}_r \rangle
=
\rho \tilde{\mathbf{l}} \tilde{u}_r .
\end{equation}
For a full derivation, we refer to Appendix~\ref{sec:favre_j}. For our
purpose, only the turbulent angular momentum flux $\mathbf{F}_{\mathrm{turb}}$
is of interest, which contains the flux carried by gravity waves and acoustic waves.
We evaluate the angular momentum flux out of region~(I) at a fixed mass coordinate of $m = 1.65 M_\odot$.
We choose a fixed mass shell rather than a fixed
radius because this mass shell stays inside the convectively stable region until around $400\, \mathrm{s}$;
a fixed analysis radius would be problematic
because of the contraction of the shells during
the simulation.

Using the numerically computed turbulent angular momentum flux,
we can now proceed to assess Equation~(\ref{eq:RWJ}).
To evaluate
the right-hand side (RHS) of Equation~(\ref{eq:e_wave}),
we use the maximum of the ratio of RMS velocity fluctuation $v_r''$ to the sound speed $c_\mathrm{s}$ in the convectively stable region containing $R_\mathrm{flux}$,

\begin{equation}
\mathcal{M}=\mathrm{max}_\mathrm{I}
\left(
\frac{(\widetilde{v_r''^2})^{1/2}}
{\tilde{c_s}}\right),
\end{equation}
and use the volume-integrated nuclear energy generation rate
$\dot{Q}_\mathrm{nuc}$ in region~(I) as a proxy for the convective luminosity
$L_\mathrm{conv}$, which is well justified under steady
state conditions \citep{Arnett2008,BM2016}\footnote{One could also define
$L_\mathrm{conv}$ as the maximum of 
the turbulent energy flux $\dot{E}$ inside a convective region.
This yields similar, but more noisy results.}.
Finally, we need the convective turnover time to evaluate Equation~(\ref{eq:RWJ}), which is computed
from the width of the convective region~(I) and
 the maximum convective velocity fluctuation $v_r''$,
\begin{equation}
 \tau = \Delta r/ \mathrm{max}_\mathrm{I} \left( v_r'' \right).
\end{equation}

\begin{figure}
 \centering
 \includegraphics[width=0.5\textwidth]{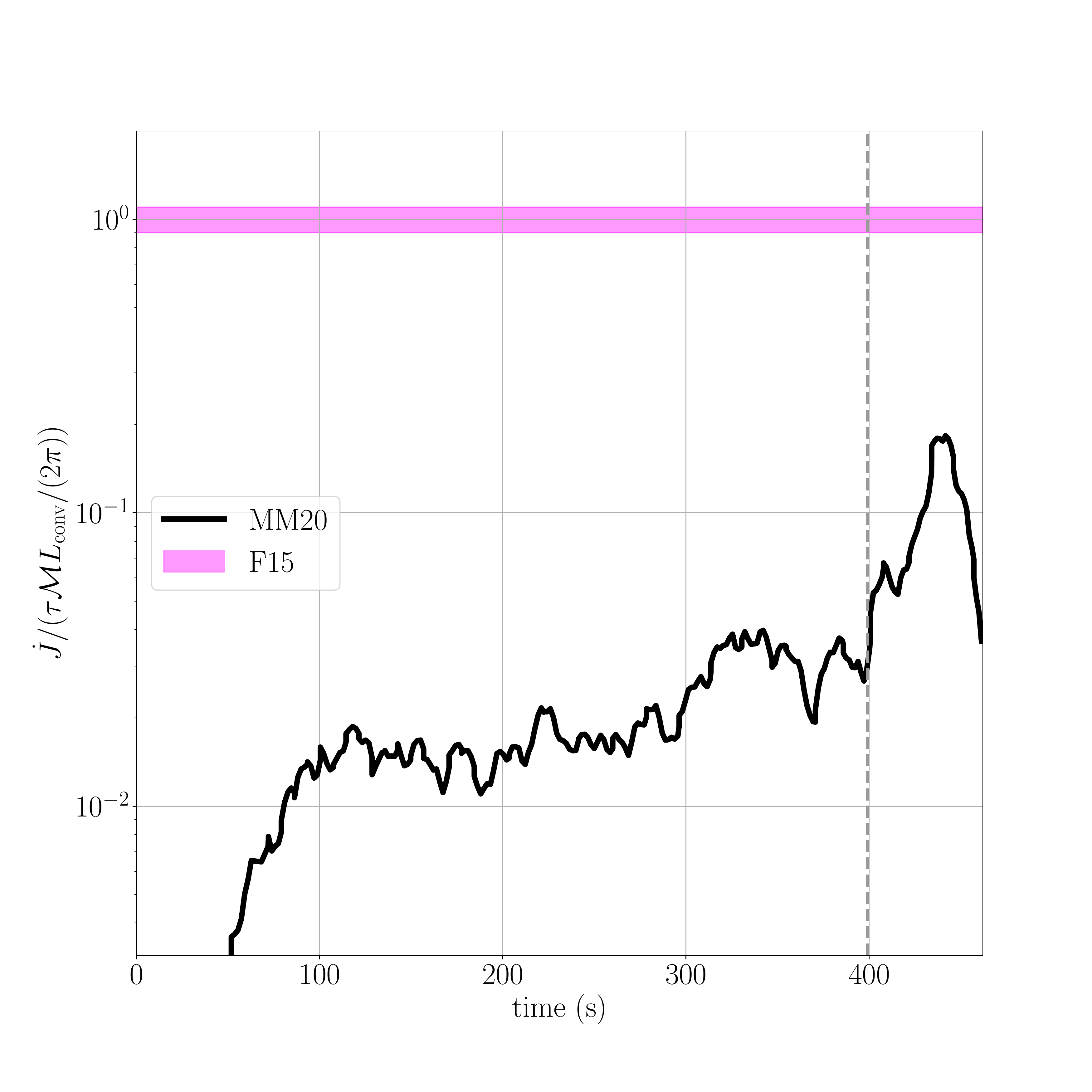}
 \caption{Ratio of the turbulent angular momentum flux $\dot{J} = |\mathbf{F}_\mathrm{turb}|$ from our 3D simulation (MM20) to the prediction 
 from Equation~(\ref{eq:RWJ}), based
 on our calculations of the convective Mach number $\mathcal{M}$ and convective burning luminosity $\dot{Q}_\mathrm{nuc}$ (and hence the energy flux via Equation~(\ref{eq:e_wave})) and convective
 turnover frequency. Our computed angular momentum flux is smaller by a factor of $\mathord{\sim}10^2$ than what the model in F15 suggests
 when using the typical wave number $\bar{m}=1$. After around $400\,\mathrm{s}$ (grey dashed line), the mass coordinate where we measure the angular
 momentum flux resides \emph{inside} the inner convective region~(I), so the ratio is no longer reliable.}
 \label{fig:Eq3scaling}
\end{figure} 

In Figure~\ref{fig:Eq3scaling}, we plot the ratio of the turbulent angular
momentum flux $\dot{J}=4\pi r^2|\mathbf{F}_{\mathrm{turb}}|$
to the value $\dot{J}_\mathrm{F15}=\tau \mathcal{M} L_\mathrm{conv}/(2\pi) $ predicted by F15.
Once convection in region~(I) has reached a steady state, the angular
momentum flux $\dot{J}$ only reaches 
 a fraction of a
few $\mathord{\sim} 10^{-2}$ of the flux
predicted by F15.
If we accept the scaling
$\dot{E}=\mathcal{M}^{\alpha} L_\mathrm{conv}$ for the wave energy
flux, 
this result implies that the effective average wave number of the
excited wave packets must be \emph{much smaller} than $\bar{m}=1$ despite the fact that the excited
modes have rather \emph{high} angular wave numbers
as discussed in Section~\ref{sec:flow}.\footnote{Alternatively, one might surmise that the excited
modes have frequencies higher than $\omega=2\pi/\tau$,
but unless there is resonant excitation it is unavoidable that
the frequency spectrum of the excited modes roughly reflects
the frequency spectrum of the turbulent driving motions.} It is, in fact, natural to expect that $\bar{m}=1$ is not a lower
bound for the average wave number if one considers the dynamics of wave
excitation at convective boundaries. Convective plumes are driven by radial
buoyancy forces and therefore hit the convective boundary almost
perpendicularly. Any overshooting plume will therefore tend to launch gravity
waves propagating in all directions away from the plume (i.e., with
wave numbers of opposing sign) with similar amplitude. A small asymmetry
between modes of different $m$ can still arise if the plume wanders around,
but the convective updrafts and downdrafts tend to be rather stationary
in simulations of convection during the neutrino-cooled burning stages, and
hence there is only a small asymmetry in amplitude between
excited modes of opposite $m$. Contrary to
the assumption of F15, $\bar{m}=1$ is therefore
not a strict lower bound for a single
wave packet launched by an overshooting plume;
$\bar{m}<1$ is easily possible and in fact
expected for waves excited by buoyancy-driven
convection.

\subsection{Testing the Random Walk Assumption}
The relatively small values of the turbulent angular momentum flux already
indicate that the mechanism for stochastic spin-up is considerably less
efficient than F15 estimated. However, there is yet another potential problem
that could further reduce the efficiency of stochastic spin-up. One might
suspect that the excitation of modes by overshooting plumes is not a
time-independent stochastic process that results in a random walk of the
angular momentum of the convective region that launches the waves, and of the
angular momentum in the destination region. Instead, there could be a
regression to zero if the stochastic excitation were to predominantly produce
prograde modes, which would drive the angular momentum of the driving region
towards zero. Such a mechanism could operate independently of
the familiar filtering mechanism for prograde and retrograde modes and
limit stochastic fluctuations of the angular momentum of convective shells
even when there is no strong differential rotation.

In order to determine whether there is such a restoring mechanism, or whether
the random walk assumption of F15 is valid, we first need to generalise
Equation~(\ref{eq:RWMa}) to accommodate the slow, secular changes in
convective luminosity and turnover time as the oxygen burning shell contracts.
We can then compare the actual time-integrated angular momentum flux to this
generalised random walk model.
It is straightforward to determine the root-mean-square
expectation value $\Delta J_\mathrm{RW}$
for a continuous random walk process by integrating
the variance of each angular momentum component
$\delta J_i^2=(\tau \dot{J}_i)^2=|\tau \dot{J}|^2/3$ of each wave packet
times the assumed correlation frequency 
of the random walk process (i.e.\ the convective
turnover frequency $\tau^{-1}$ in the
model of F15),
\begin{equation}
\Delta J_\mathrm{RW}(t) = 
\left[ 
\sum_{i=1}^{3} \int_0^t (\delta J_i)^2 \frac{\ud t'}{\tau}
 \right]^{1/2} =\left[ 
\sum_{i=1}^{3} \int_0^t \tau \dot{J}_i(t')^2 \, \ud t'
 \right]^{1/2}.
 \label{eq:J2}
\end{equation}
\begin{figure}
 \centering
 \includegraphics[width=0.5\textwidth]{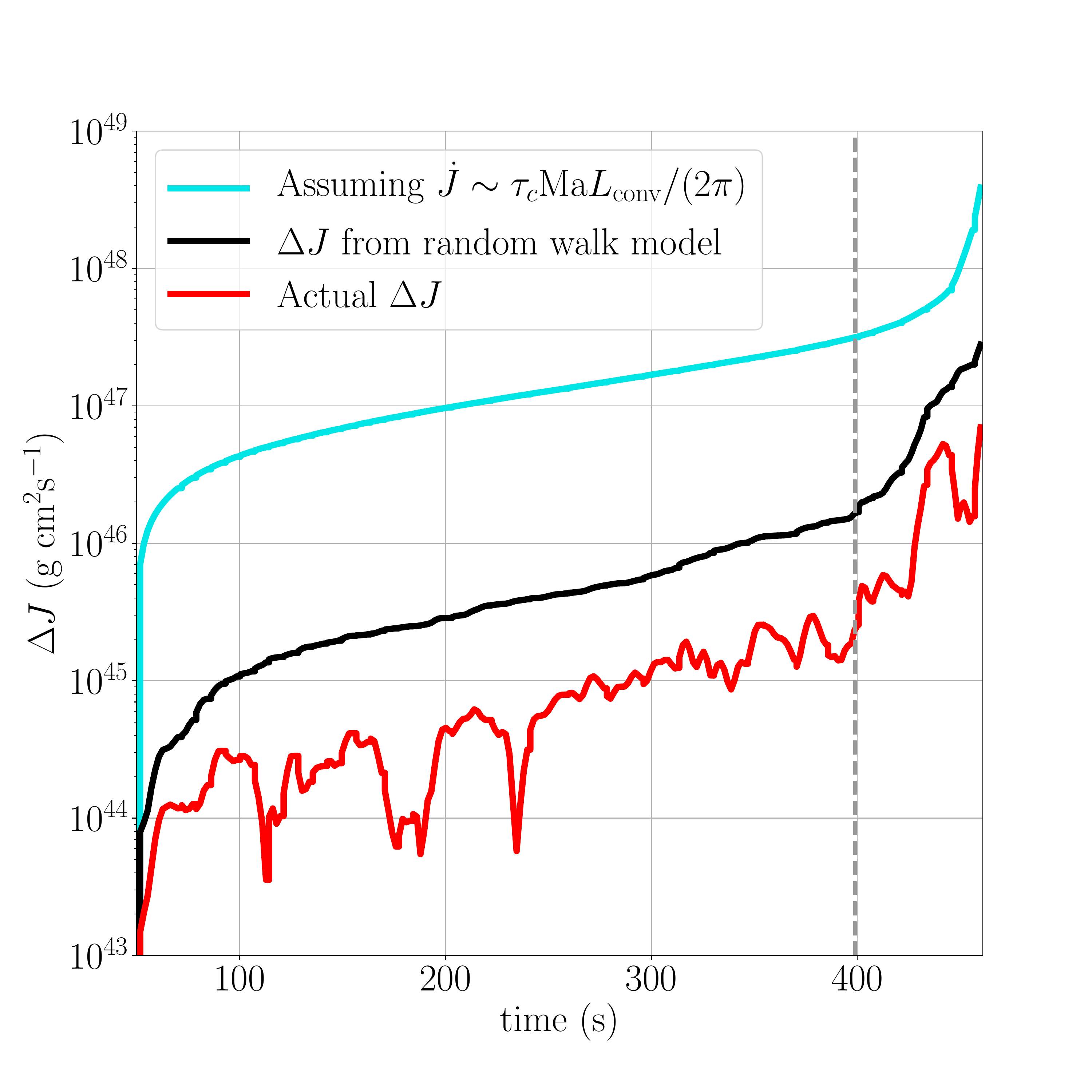}
 \caption{The change in angular momentum,
 $\Delta J$, inside the mass coordinate $m=1.65 M_\odot$
 based on the three different formulations of $\Delta J$ in Equations~(\ref{eq:J2}),(\ref{eq:J3}), and (\ref{eq:J1}). The angular momentum computed from the simulation (Equation~\ref{eq:J3}) is in red and the angular momentum computed assuming a random walk via Equation~(\ref{eq:J2}) is in black. The angular momentum calculated from
  Equation~(\ref{eq:J1}) assuming
 the relation~(\ref{eq:RWJ})
 between the angular momentum
 flux and the convective luminosity
 is in cyan. After $400\, \mathrm{s}$
 (dashed grey line) the mass coordinate where we measure the flux lies
 inside the convective region 
 (I) (see Figure~\ref{fig:earlyS}). The significant
 growth of $\Delta J$ in the last $70\,\mathrm{s}$
 is the result of stochastic angular momentum redistribution
 \emph{inside} the convective zone, and not associated with
 a flux of IGWs. The wave packet random
 walk model is no longer applicable during this phase and
 cannot be used to infer spin--up of the region from IGWs.
}
\label{fig:randomwalk2}
\end{figure} 
Figure~\ref{fig:randomwalk2} compares
$\Delta J_\mathrm{RW}$ (in black) to the actual time-integrated
angular momentum flux (in red)
$\Delta J$,
\begin{equation}
\Delta J(t)=
\left[
\sum_{i=1}^{3} 
\left(
\int_0^t \dot{J}_i(t') \, \ud t'
\right)^2
\right]^{1/2}.
\label{eq:J3}
\end{equation}
For illustrative purposes, we also show the predicted spin-up $\Delta
J_\mathrm{F15}(t)$ for the original random walk model of F15 with
$\dot{J}=\dot{E} \tau/2\pi $ (i.e.\ $\bar{m}=1$) and $\dot{E}=\mathcal{M}
L_\mathrm{conv}$ in cyan,
\begin{equation}
 \Delta J_\mathrm{F15}(t) = \frac{1}{\sqrt{3}}
 \left[
 \int_0^{t} \tau^{-1} \left(\frac{\tau^2 \mathcal{M} L_\mathrm{conv}}{2\pi}\right)^2 \, \ud t'
 \right] ^{1/2}.
 \label{eq:J1}
\end{equation}

Interestingly, even though $\Delta J_\mathrm{RW}$ (black) still overpredicts
the actual time-integrated angular momentum flux $\Delta J$ (red) by a factor
of 5 on average, this is not a huge discrepancy. 
The growth of $\Delta J$ appears roughly compatible with a random walk,
only with an effective correlation time that is $4\%$
of what is assumed in F15, since the
spin-up scales with the square root of the correlation time,
$\Delta J_\mathrm{RW}\propto \sqrt{\tau}$. This short
correlation time may be related to the rather high wave numbers
of the excited modes. The effective correlation time 
of the random walk must reflect the characteristic time scale
of the forcing motions with similar wave numbers, or in other
words, of relatively small-scale structures in the convective
flow, which evolve on significantly shorter time scales
than $\tau_\mathrm{conv}$. 

In comparing to F15, one needs to bear in mind one
important caveat on our simulation setup. What critically matters in
the model of \citet{Fuller2015} is the angular momentum flux into the convectively stable
core, while we measure the flux between two convective regions. Unlike inside the core, where
waves will damp and break, outward-propagating waves here are subject to reflection
at the next convective boundary, and it is possible that undamped waves produce standing 
waves, which potentially destructively interfere, resulting in a smaller angular momentum 
flux than the single-mode wave packet would predict. It is therefore possible that 
the discrepancy with F15 is due in part to an (incomplete) cancellation of the angular momentum flux by reflected waves, rather than to
a shorter correlation time. However, if the true correlation time of the forcing
motions were of the order of the convective turnover time $\tau_\mathrm{conv}$,
one would still expect that at least the direction of the angular momentum
carried by waves should be stable on this longer time scale.
To check this, we compute the correlation time for the direction of the
vector $\dot{J}$; a correlation time closer to the convective turnover time of
$\mathord{\sim}13\,\mathrm{s}$ would indicate that we are measuring flux from
interfering reflected waves. The correlation time of the direction of $\dot{J}$
is only $0.5 \,\mathrm{s}$, consistent with our inferred correlation time of the 
random walk. However, it remains difficult to distinguish standing and
travelling waves in Figure~\ref{fig:visit}, and we cannot rule out
some contamination of wave reflection as a contributor to the low angular
momentum. Ultimately, simulations of IGW excitation at the inner convective
boundaries will still be required to completely exclude contributions
from wave reflection.

Regardless of these potential complication, it is important to stress that
our simulation actually \emph{supports}
the assumption that stochastic wave excitation will lead to the core
angular momentum executing a random walk, albeit
with a shorter correlation time.
However, because the assumption of $\bar{m}=1$ is not valid and
because the correlation time of the random walk is somewhat
shorter than the convective turnover time in our simulations,
the original model of F15 overpredicts $\Delta J$ by
several orders of magnitude.

On the other hand, Figure~\ref{fig:randomwalk2} suggests
that the late convective burning stages might stochastically
affect the remnant angular momentum in a different manner.
During the last $70\, \mathrm{s}$ of the simulation, the
analysis radius lies \emph{inside} the oxygen shell, and
the measured $\Delta J$ indicates stochastic \emph{internal}
redistribution of angular momentum within the shell.
This stochastic redistribution of angular momentum could
potentially spin-up the inner regions of the shell
considerably. In the present example, the time-integrated
flux of angular momentum $\Delta J$ \emph{within}
the oxygen shell during the last
$70\, \mathrm{s}$ is ten times bigger than the time-integrated
wave-mediated flux of angular momentum out of the oxygen shell.
If only part of the oxygen shell is ejected during the supernova
explosion, the total angular momentum within the ``mass cut''
may be considerable. If the mass cut were placed
at $1.65 M_\odot$ (which is much larger than is realistic for this
model, cp.\ \citealp{BM2019a}), the resulting neutron star spin period would be of order $100\, \mathrm{ms}$. Naturally,
the theory of stochastic IGW excitation cannot be applied to stochastic
variations within convective regions close to the mass cut,
so one cannot make generic predictions about the importance
of this phenomenon. However, 3D supernova progenitor models
that include this effect naturally anyway are becoming more widely used.

\subsection{Estimate of neutron star spin periods}
Figure~\ref{fig:randomwalk2} shows
that during the simulation time of about 7~minutes,
the angular momentum transported out of the oxygen shell
from the outer boundary is about $2\times 10^{45}\, \mathrm{g}\,
\mathrm{cm}^2\, \mathrm{s}^{-1}$.
If a similar amount of angular momentum is transported
inwards into the core, and the angular momentum is conserved
during the collapse to a neutron star, this would
result in a neutron star rotation rate $\omega\approx 1\, \mathrm{rad}\, \mathrm{s}^{-1}$ of a typical neutron star moment of inertia
$\mathord{\sim}1.5 \times 10^{45}\, \mathrm{g}\, \mathrm{cm}^2$.
However, our simulation only covers the last phase 
in the life of the second oxygen shell. To gauge the actual impact
of stochastic spin-up, we need to extrapolate our results
to the entire lifetime of a shell similar to F15, but
with different efficiency factors.

For this purpose, it is convenient to express all relevant
quantities in Equation~(\ref{eq:RWMa}) in terms of
the shell mass $\Delta M$, inner radius $R$
and width $\Delta R$, the average $Q$-value of the nuclear
reaction per unit mass, $q$, and the nuclear time scale $\tau_\mathrm{nuc}$ for the
respective burning phase. 
Using $L_\mathrm{conv}= \Delta M\, q/\tau_\mathrm{nuc}$,
we find the convective velocity from mixing-length theory
as calibrated against 3D simulations \citep{BM2016},
\begin{equation}
v_\mathrm{conv} =
\frac{1}{2}
\left(\frac{L_\mathrm{conv} h_P} {m}\right)^{1/3}
=\frac{1}{2}\left(\frac{q R/4}{\tau_\mathrm{nuc}}\right)^{1/3},
\end{equation}
where we have taken the pressure scale height to be
$h_P=P/(\rho g) \approx R/4$, which is a good approximation
during advanced burning stages in massive stars. 
The sound speed at the bottom of the shell is approximately
\begin{equation}
c_\mathrm{s} = \left(\frac{G M}{3 R} \right)^{1/2},
\end{equation}
so that we can write the convective Mach number as
\begin{equation}
\mathcal{M} = \frac{v_\mathrm{conv}}{c_\mathrm{s}} 
= 
\frac{1}{2}
\left( \frac{q R} {4\tau_\mathrm{nuc}}\right)^{1/3} \left(\frac{G M}{3 R} \right)^{-1/2}
\label{eq:Mach},
\end{equation}
and find the convective turnover time to be $\tau_\mathrm{conv}=\Delta R/v_\mathrm{conv}$.
Putting all this together in
Equation~(\ref{eq:RWMa}) and noting that we have
 $\mathcal{N}=\tau_\mathrm{nuc}/\tau_\mathrm{conv}$
 turnovers over the lifetime of the shell, 
 we find
\begin{align}
	  \Delta J 
	  &=
	  \frac{\alpha}{\sqrt{3}} 
	  \sqrt{\frac{\tau_\mathrm{nuc}}{\tau_\mathrm{conv}}} \left(\frac{\tau_\mathrm{conv}^2 \mathcal{M} \Delta M q}{2\pi \tau_\mathrm{nuc}}\right)
	  \nonumber
	  \\
	  &\approx
	  3.4\times 10^{-3} \times \frac{\Delta R^{3/2} \Delta M q^{5/6} {R}^{1/3}}{\sqrt{G} \sqrt{M}
  \tau_\mathrm{nuc}^{1/3}},
  \label{eq:F15burn}
	\end{align}
where we have replaced the average wave number
$\bar{m}$ in Equation~(\ref{eq:RWMa}) with an efficiency
factor $\alpha\sim 10^{-2}$ in line with our simulation results.
This can be used for any of the neutrino-cooled shell burning phases,
but as pointed out by F15, one expects the last shell burning stage
to determine the stochastic spin-up at the time of collapse because
deterministic angular momentum transport can spin down the core
 again if the shell is extinguished before collapse.
 Typically, this will be the second or third oxygen shell \citep{Collins2018}, though
 in some cases, the most violent active shell at or close
 before collapse is the silicon shell as assumed by F15.
Since the last oxygen shell typically has a higher mass,
larger radius and width, shorter duration,
and a higher average $Q$-value of $q\approx 0.4\, \mathrm{MeV}/m_\mathrm{u}$
than the silicon shell, most of the factors
in Equation~(\ref{eq:F15burn}) have slightly more favorable
values than used by F15 for silicon shells.
With a neutron star moment of inertia
$I=0.36 M R_\mathrm{NS}^2$
and a neutron star
radius of $R_\mathrm{NS}=12\, \mathrm{km}$,
Equation~(\ref{eq:F15burn}) translates into a characteristic
spin period of
\begin{align}
 P_\mathrm{NS}
=&
13.5\,\mathrm{s} \times
\left(\frac{M}{1.4 M_\odot}\right)^{3/2}
\times
\left(\frac{\Delta M}{0.1 M_\odot}\right)^{-1}
\times
\left(\frac{R}{3000\, \mathrm{km}}\right)
\nonumber
\\
\label{eq:Jredq}
&
\times
\left(\frac{\Delta R}{1000\, \mathrm{km}}\right)^{-3/2}
\times
\left(\frac{\tau_\mathrm{nuc}}{10^4\, \mathrm{s}}\right)^{1/3}
\times
\left(\frac{q}{0.4\, \mathrm{MeV}/m_\mathrm{u}}\right)^{5/6},
\end{align}
in terms of typical parameters for oxygen shells in low-mass
supernova progenitors. Based on the efficiency parameters
in our simulations, we therefore expect stochastic spin-up
to play a negligible role in low-mass progenitors compared
to stochastic spin-up during during the early explosion
phase \citep{Wongwathanarat2013,BM2019a}
and by late-time fallback \citep{chan_20,stockinger_20}, which
can achieve spin periods from a few seconds down 
to milliseconds. For high-mass progenitors
with $\Delta M\sim 1 M_\odot$ and wider oxygen
shells $\Delta R \sim 5000\, \mathrm{km}$, spin periods
of order $100 \, \mathrm{ms}$ remain within reach even with
a much lower efficiency factor than assumed by F15, and stochastic
spin-up by IGWs might still be a relevant factor in determining
the neutron star spin period along with other angular momentum
processes during the progenitor evolution and spin-up/spin-down
processes during the explosion.

\section{Effects of Numerical Conservation Errors}
\label{sec:ang_cons}
Finite-volume methods as used in \textsc{Prometheus} generally cannot conserve
total angular momentum to machine accuracy (unlike smoothed-particle
hydrodynamics). Moreover, angular momentum transport in any grid-based or
particle-based simulation will be affected to some degree by numerical
dissipation. One may justifiably ask whether non-conservation of total angular
momentum or numerical dissipation have any bearing on our simulation of
stochastic spin-up. In the context of the preceding discussion, it is
especially pertinent, whether the convective region~(I) is indeed spun up in the
opposite direction to the angular momentum that is lost by waves. This
specific problem can already be addressed here, though more extensive
resolution studies and simulations with different reconstruction and
discretisation schemes naturally remain desirable for the future.

As a simple check on the quality of angular momentum conservation
in our simulation, we compare the time-integrated angular
momentum flux from region~(I) to the volume-integrated angular
momentum. Unlike when we calculated the angular momentum flux for the random walk at a fixed mass coordinate, we compute the angular momentum dissipation at a fixed radial coordinate.
so that we 
can formulate
the angular momentum ``budget'' for
region (I) using Gauss' theorem.
We define the analysis region
region C1 as the sphere contained within $r=4337\,\mathrm{km}$,
i.e.\ the boundary of C1 is in the middle of the convectively stable region between 
regions~(I) and (II) once 3D burning and convection develops.
Analytically, we have
\begin{equation}
\label{eq:int_cons}
\int\limits_\mathrm{C1} \rho \textbf{l} \, \ud V 
=
- \int\limits_0^t \oint\limits_{\pd \mathrm{C1}}
\rho \mathbf{l} \mathbf{v}\cdot \mathbf{dA}\,\ud t'.
\end{equation}

\begin{figure}
 \centering
 \includegraphics[width=0.5\textwidth]{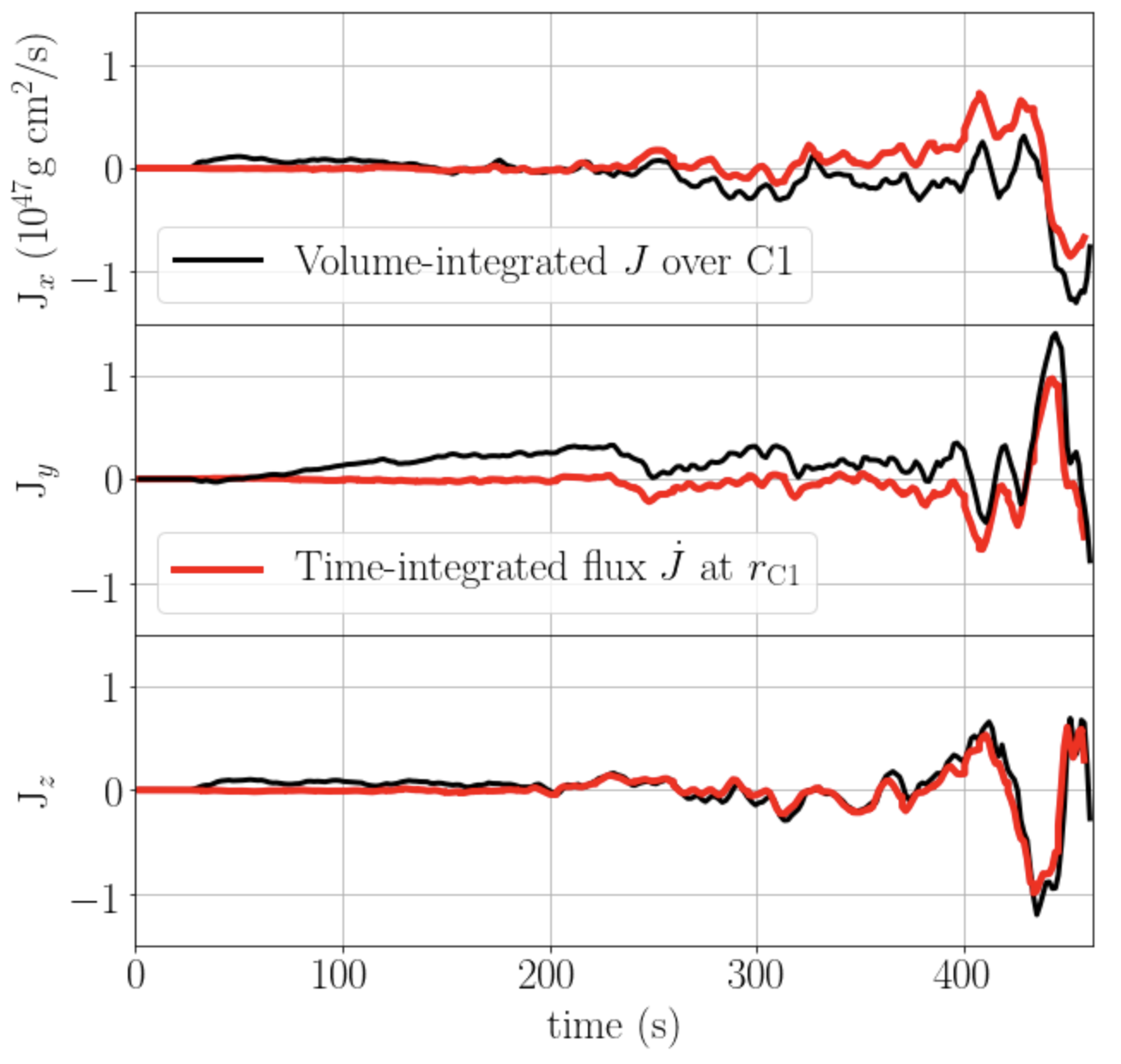}
 \caption{Time-integrated angular momentum flux at the fixed boundary C1 (red) and the (negative) integrated angular momentum over the volume enclosed by C1 (black) in the $x$, $y$ and $z$ directions. For each component, the angular momentum in the spun-up convective shell volume is consistent with the angular momentum flux leaving the volume.
 }
 \label{fig:AMxyz}
\end{figure}

In Figure~\ref{fig:AMxyz}, we compare the left-hand side and
the right-hand side of
Equation~(\ref{eq:int_cons}) for the $x$-, $y$-, and $z$-component of
the angular momentum. For all three components, at the end of the simulation the direction of the convective shell angular momentum is indeed the opposite sign of the angular momentum leaving at the flux boundary.
Early on, the evolution of $\mathbf{J}$ is dominated
by numerical errors, the $y$-component in particular
shows a significant drift over
the first $200 \, \mathrm{s}$. As soon as convection
is fully developed and the excited waves from
the convective boundary carry a significant
angular momentum flux, the 
evolution of the total angular momentum
in the analysis region clearly follows
the time-integrated angular momentum flux. This
suggests that angular momentum conservation
errors do not significantly affect our
analysis at times later than $\mathord{\sim} 200 \, \mathrm{s}$. Of course, more subtle numerical
issues, such artificial damping
of the excited waves, may limit the accuracy of our
results and need to be further investigated
in the future. However, we believe that the key
findings of this study will likely remain robust.
The code clearly has the ability to evolve relatively
weak waves, and the waves that would need to
be resolved for the mechanism of F15 to work
effectively are not of extremely short
wavelength, so that excessive dissipation is
likely not a critical problem.

\section{Conclusions}
\label{sec:conclusions}
Using the 3D hydrodynamics code \textsc{Prometheus}, we studied the angular momentum flux from stochastically excited IGWs during oxygen
shell burning in an initially non-rotating massive star. 
Our findings allow us to better assess the efficiency of
stochastic wave excitation as a process for core 
spin-up, which has been suggested by \citet{Fuller2015}
based on analytic theory.

In agreement with the theory of \citet{Fuller2015},
we find that the IGW-mediated flux of angular momentum 
out of a convective shell in a non- or slowly-rotating
progenitor can indeed be described
by a random-walk process. However we find that
the correlation time of this random-walk process 
is only a fraction of the convective turnover time
and hence significantly shorter than assumed by \citet{Fuller2015}.
This reduces the amplitude of the random walk
by a factor of several.

We also find a smaller ratio between the angular momentum flux carried by the excited IGWs and the convective luminosity 
by more than an order of magnitude than assumed by 
\citet{Fuller2015}. 
We ascribe this discrepancy to the fact that the assumption
of an average zonal wave number $\bar{m}=1$
of individual IGW ``wave packets'' is too optimistic,
since IGWs excited by overshooting
convective plumes will generally contain modes
with opposite zonal wave number $m$ of almost equal
amplitude. Therefore the average zonal wave
number $\bar{m}$ of each wave packet will be
much smaller than unity and the net angular momentum carried
by the wave packet will be very small.

The combination of a shorter effective correlation
time and a smaller instantaneous angular momentum
flux results in a time-integrated stochastic
angular momentum flux from the oxygen shell that
is about a factor of $\mathord{\sim}10^2$ smaller than predicted
by the model of \citet{Fuller2015}. It should
be pointed out, however, that our
simulation \emph{is} compatible with their analytic scaling
laws; it merely suggests that the relevant dimensionless
parameters of the model may be quite small.

The characteristic spin period from stochastic spin-up of the core by
IGWs over the entire life of a convective burning shell can be
estimated based on the mass, radius, width, and lifetime of the shell
from 1D stellar evolution models using a simple scaling relation
(Equation~\ref{eq:Jredq}).
Extrapolating our simulation results over the entire
life time of convective shells, we estimate that stochastic
spin-up by IGWs alone would result in neutron star birth
spin periods of several seconds for low-mass stars
and down to $\mathord{\sim}0.1\, \mathrm{s}$ for high-mass
stars with thick oxygen shells. These spin rates
are slower than predicted by
\citet{Fuller2015} by one to two orders of magnitude.
Thus our findings suggest that stochastic spin-up of progenitor
cores will usually not play a major role in determining
the core spin rates of massive stars and neutron
star birth periods, because stochastic spin-up processes
during the supernova explosion will impart more
angular momentum onto the neutron star.

It remains to be seen, however, to what extent the efficiency
factors for IGW excitation and stochastic core spin-up
found in our simulation depend on the detailed properties
of convective zones and the structure of the convective boundary.
For example, the correlation time of the random walk and the average
zonal wave number will depend on the typical size of the
convective eddies. Convection zones with pronounced
large-scale flow patterns and fewer but stronger plume impact
events may provide more favorable conditions for stochastic
core spin-up. In future, one should also investigate
stiffer convective boundaries with sharp entropy jumps.
The inner convective boundaries, which are of highest relevance
for the problem of stochastic spin-up, are considerably
stiffer than the convective boundary considered in this study.
Simulating IGW excitation at inner convective boundaries
would also be desirable to exclude any potential artifacts
from wave reflection at shell boundaries further outside.
Moreover, the dependence of stochastic spin-up on the convective Mach
number should be investigated numerically to confirm the
predicted scaling laws. We also note that stochastic angular
momentum redistribution \emph{within} convective zones
close to the mass cut could be relevant for predicting
neutron star birth spin periods. This effect is already
implicitly included in modern 3D supernova progenitor models.
Better understanding convection,
wave excitation, and angular momentum transport by
3D simulations will remain challenging because of
the technical difficulties of low-Mach number flow, stringent resolution
requirements at stiff convective boundaries, and the numerical problem
of angular momentum conservation. However, 3D
simulations of convective burning are clearly proving useful in
understanding angular momentum transport processes in supernova
progenitors.

\section*{Acknowledgements}
We acknowledge fruitful discussions with A.~Heger and I.~Mandel. We thank J.~Fuller, D.~Lecoanet, and the anonymous referee for helpful suggestions and comments on an earlier version which improved this manuscript. LM acknowledges support by an Australian Government Research Training (RTP) Scholarship. BM has been supported by the Australian Research Council through Future Fellowship FT160100035 and partly
as an Associate Investigator of the ARC Centre of Excllence
\emph{OzGrav} (CE170100004). This research was undertaken with the assistance of
resources from the National Computational Infrastructure
(NCI), which is supported by the Australian Government and
was supported by resources provided by the Pawsey Supercomputing
Centre with funding from the Australian Government
and the Government of Western Australia.

\section*{Data availability}
The data underlying this manuscript will be shared on reasonable request to the authors, subject to considerations of intellectual property law.



\bibliographystyle{mnras}
\bibliography{refs}    

\begin{thebibliography}{}
\makeatletter
\relax
\def\mn@urlcharsother{\let\do\@makeother \do\$\do\&\do\#\do\^\do\_\do\%\do\~}
\def\mn@doi{\begingroup\mn@urlcharsother \@ifnextchar [ {\mn@doi@}
  {\mn@doi@[]}}
\def\mn@doi@[#1]#2{\def\@tempa{#1}\ifx\@tempa\@empty \href
  {http://dx.doi.org/#2} {doi:#2}\else \href {http://dx.doi.org/#2} {#1}\fi
  \endgroup}
\def\mn@eprint#1#2{\mn@eprint@#1:#2::\@nil}
\def\mn@eprint@arXiv#1{\href {http://arxiv.org/abs/#1} {{\tt arXiv:#1}}}
\def\mn@eprint@dblp#1{\href {http://dblp.uni-trier.de/rec/bibtex/#1.xml}
  {dblp:#1}}
\def\mn@eprint@#1:#2:#3:#4\@nil{\def\@tempa {#1}\def\@tempb {#2}\def\@tempc
  {#3}\ifx \@tempc \@empty \let \@tempc \@tempb \let \@tempb \@tempa \fi \ifx
  \@tempb \@empty \def\@tempb {arXiv}\fi \@ifundefined
  {mn@eprint@\@tempb}{\@tempb:\@tempc}{\expandafter \expandafter \csname
  mn@eprint@\@tempb\endcsname \expandafter{\@tempc}}}

\bibitem[\protect\citeauthoryear{{Aerts} \& {Rogers}}{{Aerts} \&
  {Rogers}}{2015}]{Aerts2015}
{Aerts} C.,  {Rogers} T.~M.,  2015, \mn@doi [\apjl]
  {10.1088/2041-8205/806/2/L33}, \href
  {https://ui.adsabs.harvard.edu/abs/2015ApJ...806L..33A} {806, L33}

\bibitem[\protect\citeauthoryear{{Aerts} et~al.,}{{Aerts}
  et~al.}{2017}]{Aerts2017b}
{Aerts} C.,  et~al., 2017, \mn@doi [\aap] {10.1051/0004-6361/201730571}, \href
  {https://ui.adsabs.harvard.edu/abs/2017A&A...602A..32A} {602, A32}

\bibitem[\protect\citeauthoryear{{Alvan}, {Brun}  \& {Mathis}}{{Alvan}
  et~al.}{2014}]{Alvan2014}
{Alvan} L.,  {Brun} A.~S.,   {Mathis} S.,  2014, \mn@doi [\aap]
  {10.1051/0004-6361/201323253}, \href
  {https://ui.adsabs.harvard.edu/abs/2014A\&A...565A..42A} {565, A42}

\bibitem[\protect\citeauthoryear{{Alvan}, {Strugarek}, {Brun}, {Mathis}  \&
  {Garcia}}{{Alvan} et~al.}{2015}]{Alvan2015}
{Alvan} L.,  {Strugarek} A.,  {Brun} A.~S.,  {Mathis} S.,   {Garcia} R.~A.,
  2015, \mn@doi [\aap] {10.1051/0004-6361/201526250}, \href
  {https://ui.adsabs.harvard.edu/abs/2015A\&A...581A.112A} {581, A112}

\bibitem[\protect\citeauthoryear{{Andrassy}, {Herwig}, {Woodward}  \&
  {Ritter}}{{Andrassy} et~al.}{2020}]{Andrassy2018}
{Andrassy} R.,  {Herwig} F.,  {Woodward} P.,   {Ritter} C.,  2020, \mn@doi
  [\mnras] {10.1093/mnras/stz2952}, \href
  {https://ui.adsabs.harvard.edu/abs/2019MNRAS.tmp.2556A} {491, 972}

\bibitem[\protect\citeauthoryear{Arnett, Meakin  \& Young}{Arnett
  et~al.}{2008}]{Arnett2008}
Arnett D.,  Meakin C.,   Young P.~A.,  2008, \mn@doi [The Astrophysical
  Journal] {10.1088/0004-637x/690/2/1715}, 690, 1715

\bibitem[\protect\citeauthoryear{{Baldwin} et~al.,}{{Baldwin}
  et~al.}{2001}]{Baldwin2001}
{Baldwin} M.~P.,  et~al., 2001, \mn@doi [Reviews of Geophysics]
  {10.1029/1999RG000073}, \href
  {https://ui.adsabs.harvard.edu/abs/2001RvGeo..39..179B} {39, 179}

\bibitem[\protect\citeauthoryear{{Beck} et~al.,}{{Beck}
  et~al.}{2012}]{Beck2012}
{Beck} P.~G.,  et~al., 2012, \mn@doi [\nat] {10.1038/nature10612}, \href
  {https://ui.adsabs.harvard.edu/abs/2012Natur.481...55B} {481, 55}

\bibitem[\protect\citeauthoryear{{Beck} et~al.,}{{Beck}
  et~al.}{2014}]{Beck2014}
{Beck} P.~G.,  et~al., 2014, \mn@doi [\aap] {10.1051/0004-6361/201322477},
  \href {https://ui.adsabs.harvard.edu/abs/2014A\&A...564A..36B} {564, A36}

\bibitem[\protect\citeauthoryear{{Belkacem} et~al.,}{{Belkacem}
  et~al.}{2015}]{Belkacem2015}
{Belkacem} K.,  et~al., 2015, \mn@doi [\aap] {10.1051/0004-6361/201526043},
  \href {https://ui.adsabs.harvard.edu/abs/2015A\&A...579A..31B} {579, A31}

\bibitem[\protect\citeauthoryear{{Bilinski}, {Smith}, {Li}, {Williams}, {Zheng}
   \& {Filippenko}}{{Bilinski} et~al.}{2015}]{Bilinski2015}
{Bilinski} C.,  {Smith} N.,  {Li} W.,  {Williams} G.~G.,  {Zheng} W.,
  {Filippenko} A.~V.,  2015, \mn@doi [\mnras] {10.1093/mnras/stv566}, \href
  {https://ui.adsabs.harvard.edu/abs/2015MNRAS.450..246B} {450, 246}

\bibitem[\protect\citeauthoryear{{Blondin} \& {Mezzacappa}}{{Blondin} \&
  {Mezzacappa}}{2007}]{Blondin2007}
{Blondin} J.~M.,  {Mezzacappa} A.,  2007, \mn@doi [\nat] {10.1038/nature05428},
  \href {http://adsabs.harvard.edu/abs/2007Natur.445...58B} {445, 58}

\bibitem[\protect\citeauthoryear{{Cantiello}, {Mankovich}, {Bildsten},
  {Christensen-Dalsgaard}  \& {Paxton}}{{Cantiello}
  et~al.}{2014}]{Cantiello2014}
{Cantiello} M.,  {Mankovich} C.,  {Bildsten} L.,  {Christensen-Dalsgaard} J.,
  {Paxton} B.,  2014, \mn@doi [\apj] {10.1088/0004-637X/788/1/93}, \href
  {https://ui.adsabs.harvard.edu/abs/2014ApJ...788...93C} {788, 93}

\bibitem[\protect\citeauthoryear{{Chan}, {M\"uller}  \& {Heger}}{{Chan}
  et~al.}{2020}]{chan_20}
{Chan} C.,  {M\"uller} B.,   {Heger} A.,  2020, \mn@doi [Monthly Notices of the
  Royal Astronomical Society] {10.1093/mnras/staa1431}

\bibitem[\protect\citeauthoryear{{Charbonnel} \& {Talon}}{{Charbonnel} \&
  {Talon}}{2005}]{Charbonnel2005}
{Charbonnel} C.,  {Talon} S.,  2005, \mn@doi [Science]
  {10.1126/science.1116849}, \href
  {https://ui.adsabs.harvard.edu/abs/2005Sci...309.2189C} {309, 2189}

\bibitem[\protect\citeauthoryear{{Colella} \& {Sekora}}{{Colella} \&
  {Sekora}}{2008}]{colella_08}
{Colella} P.,  {Sekora} M.~D.,  2008, \mn@doi [Journal of Computational
  Physics] {10.1016/j.jcp.2008.03.034}, \href
  {http://adsabs.harvard.edu/abs/2008JCoPh.227.7069C} {227, 7069}

\bibitem[\protect\citeauthoryear{{Collins}, {M{\"u}ller}  \& {Heger}}{{Collins}
  et~al.}{2018}]{Collins2018}
{Collins} C.,  {M{\"u}ller} B.,   {Heger} A.,  2018, \mn@doi [\mnras]
  {10.1093/mnras/stx2470}, \href
  {https://ui.adsabs.harvard.edu/abs/2018MNRAS.473.1695C} {473, 1695}

\bibitem[\protect\citeauthoryear{{Cristini}, {Meakin}, {Hirschi}, {Arnett},
  {Georgy}, {Viallet}  \& {Walkington}}{{Cristini} et~al.}{2017}]{Cristini2017}
{Cristini} A.,  {Meakin} C.,  {Hirschi} R.,  {Arnett} D.,  {Georgy} C.,
  {Viallet} M.,   {Walkington} I.,  2017, \mn@doi [\mnras]
  {10.1093/mnras/stx1535}, \href
  {https://ui.adsabs.harvard.edu/abs/2017MNRAS.471..279C} {471, 279}

\bibitem[\protect\citeauthoryear{{Deheuvels} et~al.,}{{Deheuvels}
  et~al.}{2012}]{Deh2012}
{Deheuvels} S.,  et~al., 2012, \mn@doi [\apj] {10.1088/0004-637X/756/1/19},
  \href {https://ui.adsabs.harvard.edu/abs/2012ApJ...756...19D} {756, 19}

\bibitem[\protect\citeauthoryear{{Deheuvels} et~al.,}{{Deheuvels}
  et~al.}{2014}]{Deh2014}
{Deheuvels} S.,  et~al., 2014, \mn@doi [\aap] {10.1051/0004-6361/201322779},
  \href {https://ui.adsabs.harvard.edu/abs/2014A\&A...564A..27D} {564, A27}

\bibitem[\protect\citeauthoryear{{Edelmann}, {Ratnasingam}, {Pedersen},
  {Bowman}, {Prat}  \& {Rogers}}{{Edelmann} et~al.}{2019}]{Edelmann2019}
{Edelmann} P.~V.~F.,  {Ratnasingam} R.~P.,  {Pedersen} M.~G.,  {Bowman} D.~M.,
  {Prat} V.,   {Rogers} T.~M.,  2019, \mn@doi [\apj]
  {10.3847/1538-4357/ab12df}, \href
  {https://ui.adsabs.harvard.edu/abs/2019ApJ...876....4E} {876, 4}

\bibitem[\protect\citeauthoryear{{Faucher-Gigu{\`e}re} \&
  {Kaspi}}{{Faucher-Gigu{\`e}re} \& {Kaspi}}{2006}]{Fau2006}
{Faucher-Gigu{\`e}re} C.-A.,  {Kaspi} V.~M.,  2006, \mn@doi [\apj]
  {10.1086/501516}, \href
  {https://ui.adsabs.harvard.edu/abs/2006ApJ...643..332F} {643, 332}

\bibitem[\protect\citeauthoryear{{Favre}}{{Favre}}{1965}]{Favre1965}
{Favre} A.~J.,  1965, \mn@doi [Journal of Applied Mechanics]
  {10.1115/1.3625792}, \href
  {https://ui.adsabs.harvard.edu/abs/1965JAM....32..241F} {32, 241}

\bibitem[\protect\citeauthoryear{{Foley}, {Berger}, {Fox}, {Levesque},
  {Challis}, {Ivans}, {Rhoads}  \& {Soderberg}}{{Foley}
  et~al.}{2011}]{Foley2011}
{Foley} R.~J.,  {Berger} E.,  {Fox} O.,  {Levesque} E.~M.,  {Challis} P.~J.,
  {Ivans} I.~I.,  {Rhoads} J.~E.,   {Soderberg} A.~M.,  2011, \mn@doi [\apj]
  {10.1088/0004-637X/732/1/32}, \href
  {https://ui.adsabs.harvard.edu/abs/2011ApJ...732...32F} {732, 32}

\bibitem[\protect\citeauthoryear{{Fryxell}, {M\"uller}  \& {Arnett}}{{Fryxell}
  et~al.}{1989}]{fryxell_89}
{Fryxell} B.~A.,  {M\"uller} E.,   {Arnett} D.,  1989, Max-Planck-Institut
  f\"ur Astrophysik, Preprint0, 449

\bibitem[\protect\citeauthoryear{{Fryxell}, {Arnett}  \& {Mueller}}{{Fryxell}
  et~al.}{1991}]{fryxell_91}
{Fryxell} B.,  {Arnett} D.,   {Mueller} E.,  1991, \mn@doi [\apj]
  {10.1086/169657}, \href {http://adsabs.harvard.edu/abs/1991ApJ...367..619F}
  {367, 619}

\bibitem[\protect\citeauthoryear{{Fuller}}{{Fuller}}{2017}]{Fuller2017}
{Fuller} J.,  2017, \mn@doi [\mnras] {10.1093/mnras/stx1314}, \href
  {https://ui.adsabs.harvard.edu/\#abs/2017MNRAS.470.1642F} {470, 1642}

\bibitem[\protect\citeauthoryear{Fuller, Lecoanet, Cantiello  \& Brown}{Fuller
  et~al.}{2014}]{Fuller2014}
Fuller J.,  Lecoanet D.,  Cantiello M.,   Brown B.,  2014, \mn@doi [The
  Astrophysical Journal] {10.1088/0004-637x/796/1/17}, 796, 17

\bibitem[\protect\citeauthoryear{{Fuller}, {Cantiello}, {Lecoanet}  \&
  {Quataert}}{{Fuller} et~al.}{2015}]{Fuller2015}
{Fuller} J.,  {Cantiello} M.,  {Lecoanet} D.,   {Quataert} E.,  2015, \mn@doi
  [\apj] {10.1088/0004-637X/810/2/101}, \href
  {https://ui.adsabs.harvard.edu/\#abs/2015ApJ...810..101F} {810, 101}

\bibitem[\protect\citeauthoryear{{Fuller}, {Piro}  \& {Jermyn}}{{Fuller}
  et~al.}{2019}]{Fuller2019}
{Fuller} J.,  {Piro} A.~L.,   {Jermyn} A.~S.,  2019, \mn@doi [\mnras]
  {10.1093/mnras/stz514}, \href
  {https://ui.adsabs.harvard.edu/abs/2019MNRAS.485.3661F} {485, 3661}

\bibitem[\protect\citeauthoryear{{Goldreich} \& {Kumar}}{{Goldreich} \&
  {Kumar}}{1990}]{GK1990}
{Goldreich} P.,  {Kumar} P.,  1990, \mn@doi [\apj] {10.1086/169376}, \href
  {http://adsabs.harvard.edu/abs/1990ApJ...363..694G} {363, 694}

\bibitem[\protect\citeauthoryear{{Gull{\'o}n}, {Miralles}, {Vigan{\`o}}  \&
  {Pons}}{{Gull{\'o}n} et~al.}{2014}]{Gullon2014}
{Gull{\'o}n} M.,  {Miralles} J.~A.,  {Vigan{\`o}} D.,   {Pons} J.~A.,  2014,
  \mn@doi [\mnras] {10.1093/mnras/stu1253}, \href
  {https://ui.adsabs.harvard.edu/abs/2014MNRAS.443.1891G} {443, 1891}

\bibitem[\protect\citeauthoryear{{Heger} \& {Woosley}}{{Heger} \&
  {Woosley}}{2010}]{heger_10}
{Heger} A.,  {Woosley} S.~E.,  2010, \mn@doi [\apj]
  {10.1088/0004-637X/724/1/341}, \href
  {http://adsabs.harvard.edu/abs/2010ApJ...724..341H} {724, 341}

\bibitem[\protect\citeauthoryear{{Heger}, {Langer}  \& {Woosley}}{{Heger}
  et~al.}{2000}]{Heger2000}
{Heger} A.,  {Langer} N.,   {Woosley} S.~E.,  2000, \mn@doi [\apj]
  {10.1086/308158}, \href
  {https://ui.adsabs.harvard.edu/abs/2000ApJ...528..368H} {528, 368}

\bibitem[\protect\citeauthoryear{{Heger}, {Woosley}  \& {Spruit}}{{Heger}
  et~al.}{2005}]{Heger2005}
{Heger} A.,  {Woosley} S.~E.,   {Spruit} H.~C.,  2005, \mn@doi [\apj]
  {10.1086/429868}, \href
  {https://ui.adsabs.harvard.edu/abs/2005ApJ...626..350H} {626, 350}

\bibitem[\protect\citeauthoryear{{Hirschi}, {Meynet}  \& {Maeder}}{{Hirschi}
  et~al.}{2004}]{hirschi_04}
{Hirschi} R.,  {Meynet} G.,   {Maeder} A.,  2004, \mn@doi [\aap]
  {10.1051/0004-6361:20041095}, \href
  {https://ui.adsabs.harvard.edu/abs/2004A&A...425..649H} {425, 649}

\bibitem[\protect\citeauthoryear{{Horst}, {Edelmann}, {Andrassy}, {Roepke},
  {Bowman}, {Aerts}  \& {Ratnasingam}}{{Horst} et~al.}{2020}]{Horst2020}
{Horst} L.,  {Edelmann} P.~V.~F.,  {Andrassy} R.,  {Roepke} F.~K.,  {Bowman}
  D.~M.,  {Aerts} C.,   {Ratnasingam} R.~P.,  2020, arXiv e-prints, \href
  {https://ui.adsabs.harvard.edu/abs/2020arXiv200603011H} {p. arXiv:2006.03011}

\bibitem[\protect\citeauthoryear{{Jones}, {Andrassy}, {Sandalski}, {Davis},
  {Woodward}  \& {Herwig}}{{Jones} et~al.}{2017}]{Jones2017}
{Jones} S.,  {Andrassy} R.,  {Sandalski} S.,  {Davis} A.,  {Woodward} P.,
  {Herwig} F.,  2017, \mn@doi [\mnras] {10.1093/mnras/stw2783}, \href
  {https://ui.adsabs.harvard.edu/abs/2017MNRAS.465.2991J} {465, 2991}

\bibitem[\protect\citeauthoryear{Kageyama \& Sato}{Kageyama \&
  Sato}{2004}]{kageyama_04}
Kageyama A.,  Sato T.,  2004, \mn@doi [Geochemistry, Geophysics, Geosystems]
  {10.1029/2004GC000734}, 5, n/a

\bibitem[\protect\citeauthoryear{{Kazeroni}, {Guilet}  \&
  {Foglizzo}}{{Kazeroni} et~al.}{2016}]{Kazeroni2016}
{Kazeroni} R.,  {Guilet} J.,   {Foglizzo} T.,  2016, \mn@doi [\mnras]
  {10.1093/mnras/stv2666}, \href
  {http://adsabs.harvard.edu/abs/2016MNRAS.456..126K} {456, 126}

\bibitem[\protect\citeauthoryear{{Kim} \& {MacGregor}}{{Kim} \&
  {MacGregor}}{2001}]{Kim2001}
{Kim} E.-j.,  {MacGregor} K.~B.,  2001, \mn@doi [\apjl] {10.1086/322973}, \href
  {https://ui.adsabs.harvard.edu/abs/2001ApJ...556L.117K} {556, L117}

\bibitem[\protect\citeauthoryear{{Kumar}, {Talon}  \& {Zahn}}{{Kumar}
  et~al.}{1999}]{Kumar1999}
{Kumar} P.,  {Talon} S.,   {Zahn} J.-P.,  1999, \mn@doi [\apj]
  {10.1086/307464}, \href
  {https://ui.adsabs.harvard.edu/abs/1999ApJ...520..859K} {520, 859}

\bibitem[\protect\citeauthoryear{{Lighthill}}{{Lighthill}}{1952}]{Lighthill1952}
{Lighthill} M.~J.,  1952, \mn@doi [Proceedings of the Royal Society of London
  Series A] {10.1098/rspa.1952.0060}, \href
  {https://ui.adsabs.harvard.edu/abs/1952RSPSA.211..564L} {211, 564}

\bibitem[\protect\citeauthoryear{{Marshall}, {Gotthelf}, {Zhang}, {Middleditch}
   \& {Wang}}{{Marshall} et~al.}{1998}]{Marshall1998}
{Marshall} F.~E.,  {Gotthelf} E.~V.,  {Zhang} W.,  {Middleditch} J.,   {Wang}
  Q.~D.,  1998, \mn@doi [\apjl] {10.1086/311381}, \href
  {https://ui.adsabs.harvard.edu/abs/1998ApJ...499L.179M} {499, L179}

\bibitem[\protect\citeauthoryear{Mcley \& Soker}{Mcley \&
  Soker}{2014}]{Mcley2014}
Mcley L.,  Soker N.,  2014, \mn@doi [Monthly Notices of the Royal Astronomical
  Society] {10.1093/mnras/stu1952}, 445, 2492

\bibitem[\protect\citeauthoryear{Meakin \& Arnett}{Meakin \&
  Arnett}{2007}]{Meakin2007}
Meakin C.~A.,  Arnett D.,  2007, \mn@doi [The Astrophysical Journal]
  {10.1086/520318}, 667, 448

\bibitem[\protect\citeauthoryear{{Melson}, {Janka}  \& {Marek}}{{Melson}
  et~al.}{2015}]{melson_15a}
{Melson} T.,  {Janka} H.-T.,   {Marek} A.,  2015, \mn@doi [\apjl]
  {10.1088/2041-8205/801/2/L24}, \href
  {http://adsabs.harvard.edu/abs/2015ApJ...801L..24M} {801, L24}

\bibitem[\protect\citeauthoryear{{Moc{\'a}k}, {M{\"u}ller}, {Weiss}  \&
  {Kifonidis}}{{Moc{\'a}k} et~al.}{2008}]{Mocak2008}
{Moc{\'a}k} M.,  {M{\"u}ller} E.,  {Weiss} A.,   {Kifonidis} K.,  2008, \mn@doi
  [\aap] {10.1051/0004-6361:200810169}, \href
  {https://ui.adsabs.harvard.edu/abs/2008A&A...490..265M} {490, 265}

\bibitem[\protect\citeauthoryear{{Mosser} et~al.,}{{Mosser}
  et~al.}{2012}]{Mosser2012}
{Mosser} B.,  et~al., 2012, \mn@doi [\aap] {10.1051/0004-6361/201220106}, \href
  {https://ui.adsabs.harvard.edu/abs/2012A\&A...548A..10M} {548, A10}

\bibitem[\protect\citeauthoryear{{M{\"u}ller}}{{M{\"u}ller}}{2019}]{BM2019b}
{M{\"u}ller} B.,  2019, \mn@doi [\mnras] {10.1093/mnras/stz1594}, \href
  {https://ui.adsabs.harvard.edu/abs/2019MNRAS.487.5304M} {487, 5304}

\bibitem[\protect\citeauthoryear{{M{\"u}ller}, {Viallet}, {Heger}  \&
  {Janka}}{{M{\"u}ller} et~al.}{2016}]{BM2016}
{M{\"u}ller} B.,  {Viallet} M.,  {Heger} A.,   {Janka} H.-T.,  2016, \mn@doi
  [\apj] {10.3847/1538-4357/833/1/124}, \href
  {https://ui.adsabs.harvard.edu/abs/2016ApJ...833..124M} {833, 124}

\bibitem[\protect\citeauthoryear{{M{\"u}ller} et~al.,}{{M{\"u}ller}
  et~al.}{2019}]{BM2019a}
{M{\"u}ller} B.,  et~al., 2019, \mn@doi [\mnras] {10.1093/mnras/stz216}, \href
  {http://adsabs.harvard.edu/abs/2019MNRAS.484.3307M} {484, 3307}

\bibitem[\protect\citeauthoryear{{Noutsos}, {Schnitzeler}, {Keane}, {Kramer}
  \& {Johnston}}{{Noutsos} et~al.}{2013}]{Noutsos2013}
{Noutsos} A.,  {Schnitzeler} D.~H.~F.~M.,  {Keane} E.~F.,  {Kramer} M.,
  {Johnston} S.,  2013, \mn@doi [\mnras] {10.1093/mnras/stt047}, \href
  {https://ui.adsabs.harvard.edu/abs/2013MNRAS.430.2281N} {430, 2281}

\bibitem[\protect\citeauthoryear{{Pin{\c{c}}on}, {Belkacem}  \&
  {Goupil}}{{Pin{\c{c}}on} et~al.}{2016}]{Pincon2016}
{Pin{\c{c}}on} C.,  {Belkacem} K.,   {Goupil} M.~J.,  2016, \mn@doi [\aap]
  {10.1051/0004-6361/201527663}, \href
  {https://ui.adsabs.harvard.edu/abs/2016A\&A...588A.122P} {588, A122}

\bibitem[\protect\citeauthoryear{{Pin{\c{c}}on}, {Belkacem}, {Goupil}  \&
  {Marques}}{{Pin{\c{c}}on} et~al.}{2017}]{Pincon2017}
{Pin{\c{c}}on} C.,  {Belkacem} K.,  {Goupil} M.~J.,   {Marques} J.~P.,  2017,
  \mn@doi [\aap] {10.1051/0004-6361/201730998}, \href
  {https://ui.adsabs.harvard.edu/abs/2017A\&A...605A..31P} {605, A31}

\bibitem[\protect\citeauthoryear{Pope}{Pope}{2000}]{Pope2000}
Pope S.~B.,  2000, Turbulent Flows.
Cambridge University Press, \mn@doi{10.1017/CBO9780511840531}

\bibitem[\protect\citeauthoryear{Popov \& Turolla}{Popov \&
  Turolla}{2012}]{Popov2012}
Popov S.~B.,  Turolla R.,  2012, \mn@doi [Monthly Notices of the Royal
  Astronomical Society: Letters] {10.1111/j.1745-3933.2012.01220.x}, 421, L127

\bibitem[\protect\citeauthoryear{{Popov}, {Pons}, {Miralles}, {Boldin}  \&
  {Posselt}}{{Popov} et~al.}{2010}]{Popov2010}
{Popov} S.~B.,  {Pons} J.~A.,  {Miralles} J.~A.,  {Boldin} P.~A.,   {Posselt}
  B.,  2010, \mn@doi [\mnras] {10.1111/j.1365-2966.2009.15850.x}, \href
  {https://ui.adsabs.harvard.edu/abs/2010MNRAS.401.2675P} {401, 2675}

\bibitem[\protect\citeauthoryear{Quataert \& Lecoanet}{Quataert \&
  Lecoanet}{2013}]{LQ2013}
Quataert E.,  Lecoanet D.,  2013, \mn@doi [Monthly Notices of the Royal
  Astronomical Society] {10.1093/mnras/stt055}, 430, 2363

\bibitem[\protect\citeauthoryear{{Quataert} \& {Shiode}}{{Quataert} \&
  {Shiode}}{2012}]{QS2012}
{Quataert} E.,  {Shiode} J.,  2012, \mn@doi [\mnras]
  {10.1111/j.1745-3933.2012.01264.x}, \href
  {https://ui.adsabs.harvard.edu/\#abs/2012MNRAS.423L..92Q} {423, L92}

\bibitem[\protect\citeauthoryear{{Rantsiou}, {Burrows}, {Nordhaus}  \&
  {Almgren}}{{Rantsiou} et~al.}{2011}]{Rantsiou2011}
{Rantsiou} E.,  {Burrows} A.,  {Nordhaus} J.,   {Almgren} A.,  2011, \mn@doi
  [\apj] {10.1088/0004-637X/732/1/57}, \href
  {http://adsabs.harvard.edu/abs/2011ApJ...732...57R} {732, 57}

\bibitem[\protect\citeauthoryear{{Rogers} \& {Glatzmaier}}{{Rogers} \&
  {Glatzmaier}}{2006}]{Rogers2006}
{Rogers} T.~M.,  {Glatzmaier} G.~A.,  2006, \mn@doi [\apj] {10.1086/507259},
  \href {https://ui.adsabs.harvard.edu/abs/2006ApJ...653..756R} {653, 756}

\bibitem[\protect\citeauthoryear{Rogers, Lin, McElwaine  \& Lau}{Rogers
  et~al.}{2013}]{Rogers2013}
Rogers T.~M.,  Lin D. N.~C.,  McElwaine J.~N.,   Lau H. H.~B.,  2013, \mn@doi
  [The Astrophysical Journal] {10.1088/0004-637x/772/1/21}, 772, 21

\bibitem[\protect\citeauthoryear{{Shu}}{{Shu}}{1992}]{shu}
{Shu} F.~H.,  1992, {Physics of Astrophysics, Vol. II}.
University Science Books

\bibitem[\protect\citeauthoryear{{Smith}}{{Smith}}{2014}]{Smith2014}
{Smith} N.,  2014, \mn@doi [\araa] {10.1146/annurev-astro-081913-040025}, \href
  {https://ui.adsabs.harvard.edu/abs/2014ARA\&A..52..487S} {52, 487}

\bibitem[\protect\citeauthoryear{Smith}{Smith}{2017}]{Smith2017}
Smith N.,  2017, Interacting Supernovae: Types IIn and Ibn.
Springer International Publishing, Cham, pp 403--429,
  \mn@doi{10.1007/978-3-319-21846-5_38}, \url
  {https://doi.org/10.1007/978-3-319-21846-5_38}

\bibitem[\protect\citeauthoryear{{Spruit}}{{Spruit}}{2002}]{Spruit2002}
{Spruit} H.~C.,  2002, \mn@doi [\aap] {10.1051/0004-6361:20011465}, 381, 923

\bibitem[\protect\citeauthoryear{{Stockinger} et~al.,}{{Stockinger}
  et~al.}{2020}]{stockinger_20}
{Stockinger} G.,  et~al., 2020, arXiv e-prints, \href
  {https://ui.adsabs.harvard.edu/abs/2020arXiv200502420S} {p. arXiv:2005.02420}

\bibitem[\protect\citeauthoryear{{Suijs}, {Langer}, {Poelarends}, {Yoon},
  {Heger}  \& {Herwig}}{{Suijs} et~al.}{2008}]{Suijs2008}
{Suijs} M.~P.~L.,  {Langer} N.,  {Poelarends} A.~J.,  {Yoon} S.~C.,  {Heger}
  A.,   {Herwig} F.,  2008, \mn@doi [\aap] {10.1051/0004-6361:200809411}, \href
  {https://ui.adsabs.harvard.edu/abs/2008A&A...481L..87S} {481, L87}

\bibitem[\protect\citeauthoryear{{Talon} \& {Charbonnel}}{{Talon} \&
  {Charbonnel}}{2003}]{Talon2003}
{Talon} S.,  {Charbonnel} C.,  2003, \mn@doi [\aap]
  {10.1051/0004-6361:20030672}, \href
  {https://ui.adsabs.harvard.edu/abs/2003A\&A...405.1025T} {405, 1025}

\bibitem[\protect\citeauthoryear{{Talon} \& {Charbonnel}}{{Talon} \&
  {Charbonnel}}{2005}]{Talon2005}
{Talon} S.,  {Charbonnel} C.,  2005, \mn@doi [\aap]
  {10.1051/0004-6361:20053020}, \href
  {https://ui.adsabs.harvard.edu/abs/2005A&A...440..981T} {440, 981}

\bibitem[\protect\citeauthoryear{{Talon}, {Kumar}  \& {Zahn}}{{Talon}
  et~al.}{2002}]{Talon2002}
{Talon} S.,  {Kumar} P.,   {Zahn} J.-P.,  2002, \mn@doi [\apjl]
  {10.1086/342526}, \href
  {https://ui.adsabs.harvard.edu/abs/2002ApJ...574L.175T} {574, L175}

\bibitem[\protect\citeauthoryear{{Theureau} et~al.,}{{Theureau}
  et~al.}{2011}]{Theureau2011}
{Theureau} G.,  et~al., 2011, \mn@doi [\aap] {10.1051/0004-6361/201015317},
  \href {https://ui.adsabs.harvard.edu/abs/2011A&A...525A..94T} {525, A94}

\bibitem[\protect\citeauthoryear{{Townsend}}{{Townsend}}{1966}]{Townsend1966}
{Townsend} A.~A.,  1966, \mn@doi [Journal of Fluid Mechanics]
  {10.1017/S0022112066000661}, \href
  {https://ui.adsabs.harvard.edu/abs/1966JFM....24..307T} {24, 307}

\bibitem[\protect\citeauthoryear{{Weaver}, {Zimmerman}  \& {Woosley}}{{Weaver}
  et~al.}{1978}]{weaver_78}
{Weaver} T.~A.,  {Zimmerman} G.~B.,   {Woosley} S.~E.,  1978, \mn@doi [\apj]
  {10.1086/156569}, \href {http://adsabs.harvard.edu/abs/1978ApJ...225.1021W}
  {225, 1021}

\bibitem[\protect\citeauthoryear{{Wheeler}, {Kagan}  \&
  {Chatzopoulos}}{{Wheeler} et~al.}{2015}]{Wheeler2015}
{Wheeler} J.~C.,  {Kagan} D.,   {Chatzopoulos} E.,  2015, \mn@doi [\apj]
  {10.1088/0004-637X/799/1/85}, \href
  {https://ui.adsabs.harvard.edu/abs/2015ApJ...799...85W} {799, 85}

\bibitem[\protect\citeauthoryear{{Wongwathanarat}, {Janka}  \&
  {M{\"u}ller}}{{Wongwathanarat} et~al.}{2013}]{Wongwathanarat2013}
{Wongwathanarat} A.,  {Janka} H.-T.,   {M{\"u}ller} E.,  2013, \mn@doi [\aap]
  {10.1051/0004-6361/201220636}, \href
  {http://adsabs.harvard.edu/abs/2013A\%26A...552A.126W} {552, A126}

\bibitem[\protect\citeauthoryear{{Woosley} \& {Heger}}{{Woosley} \&
  {Heger}}{2015}]{Woosley2015}
{Woosley} S.~E.,  {Heger} A.,  2015, \mn@doi [\apj]
  {10.1088/0004-637X/810/1/34}, \href
  {http://adsabs.harvard.edu/abs/2015ApJ...810...34W} {810, 34}

\bibitem[\protect\citeauthoryear{{Woosley}, {Blinnikov}  \& {Heger}}{{Woosley}
  et~al.}{2007}]{Woosley2007}
{Woosley} S.~E.,  {Blinnikov} S.,   {Heger} A.,  2007, \mn@doi [\nat]
  {10.1038/nature06333}, \href
  {https://ui.adsabs.harvard.edu/abs/2007Natur.450..390W} {450, 390}

\bibitem[\protect\citeauthoryear{{Yadav}, {M{\"u}ller}, {Janka}, {Melson}  \&
  {Heger}}{{Yadav} et~al.}{2020}]{Yadav2019}
{Yadav} N.,  {M{\"u}ller} B.,  {Janka} H.~T.,  {Melson} T.,   {Heger} A.,
  2020, \mn@doi [\apj] {10.3847/1538-4357/ab66bb}, \href
  {https://ui.adsabs.harvard.edu/abs/2020ApJ...890...94Y} {890, 94}

\bibitem[\protect\citeauthoryear{{Zahn}, {Talon}  \& {Matias}}{{Zahn}
  et~al.}{1997}]{Zahn1997}
{Zahn} J.~P.,  {Talon} S.,   {Matias} J.,  1997, \aap, \href
  {https://ui.adsabs.harvard.edu/abs/1997A\&A...322..320Z} {322, 320}

\bibitem[\protect\citeauthoryear{{den Hartogh}, {Eggenberger}  \&
  {Deheuvels}}{{den Hartogh} et~al.}{2020}]{Denhartough2020}
{den Hartogh} J.~W.,  {Eggenberger} P.,   {Deheuvels} S.,  2020, \mn@doi [\aap]
  {10.1051/0004-6361/202037568}, \href
  {https://ui.adsabs.harvard.edu/abs/2020A&A...634L..16D} {634, L16}

\makeatother
\end{thebibliography}



\onecolumn
\appendix
\section{Favre-averaged Angular Momentum Transport Equation} 
\label{sec:favre_j}
To derive the Favre-averaged angular momentum transport equation, we
cross $\mathbf{r}$ with the momentum equation \citep{shu,Pope2000},
\begin{equation}
\mathbf{r} \times 
\left [ \frac{\partial (\rho \mathbf{u})}{\partial t} 
+ \nabla \cdot (\rho \mathbf{u} \otimes \mathbf{u})\right] 
\nabla P+
= \mathbf{r} \times \rho \mathbf{g}.
\end{equation}
It is convenient to write this in component form as,
\begin{equation}
 \frac{\partial (\epsilon_{ijk} r_j \rho u_k)}{\partial t} 
+ \epsilon_{ijk} r_j \nabla_l (\rho u_l u_k +\delta_{lk} P)
= 
\epsilon_{ijk} r_j \rho g_k,
\end{equation}
which leads to
\begin{equation}
 \frac{\partial (\epsilon_{ijk} r_j \rho u_k)}{\partial t} 
+ \nabla_l [\epsilon_{ijk} r_j (\rho u_l  u_k + P \delta_{lk})]
- \epsilon_{ijk} 
(\rho u_l u_k+ P \delta_{lk}) \underbrace{\nabla_l r_j}_{=\delta_{lj}}
= 
\epsilon_{ijk} r_j \rho g_k,
\end{equation}
and hence
\begin{equation}
 \frac{\partial (\epsilon_{ijk} r_j \rho u_k)}{\partial t} 
+ \nabla_l [\epsilon_{ijk} r_j ( \rho u_l  u_k+ P \delta_{lk})]
- \underbrace{\epsilon_{ijk} (\rho u_j u_k+P \delta_{lk}) }_{=0}
= 
\epsilon_{ijk} r_j \rho g_k.
\end{equation}
This can again be written in component-free notation
\begin{equation}
 \frac{\partial \rho \mathbf{l}}{\partial t} 
+ \nabla \cdot (\rho \mathbf{u}\otimes \mathbf{l})
+\nabla \cdot (*\mathbf{r} P )
=
\mathbf{r}\times \rho \mathbf{g},
\end{equation}
where $\mathbf{l}=\mathbf{r}\times\mathbf{v}$ is the specific
angular momentum and $*$ denotes the Hodge star operator.
The cross product on the right-hand side vanishes if we assume
a monopole potential, and the second divergence term
on the left-hand side vanishes when we average over
a thin spherical shell
\begin{equation}
\int \nabla \cdot (*\mathbf{r} P ) \,\ud V
=
\int_{\pd V} P \mathbf{r} \times \mathbf{dA}
=
\int_{\pd V} P \mathbf{r} \times \mathbf{n} \,\ud A
=0.
\end{equation}
Hence we need only take into account the first two
terms on the left-hand side when performing a spherical
Favre decomposition. Decomposing the velocity
and averaging over spherical surfaces yields
\begin{equation}
 \frac{\partial \langle \rho \mathbf{l}\rangle}{\partial t} 
+ \nabla_r \cdot
\langle
\rho
(\tilde{u}_r+u_r'')
(\mathbf{\tilde{l}}+\mathbf{l''})
\rangle
= 0,
\end{equation}
where $\nabla_r$ denotes the radial part of the divergence operator.
Using the usual rules 
$\langle\rho \tilde{X} \tilde{Y}\rangle=
\hat{\rho} \tilde{X} \tilde{Y}$
and
$\langle\rho \tilde{X} Y''\rangle=0$
for correlation terms containing no or only one
fluctuating intensive quantity, only two terms remain,
\begin{equation}
 \frac{\partial \langle \rho \mathbf{l}\rangle}{\partial t} 
+ \nabla \cdot 
(
\hat{\rho}
\tilde{u}_r
\mathbf{\tilde{l}}
)
+ \nabla \cdot 
\langle
\rho
u_r''
\mathbf{l''}
\rangle
= 0.
\end{equation}

\bsp	
\label{lastpage}
\end{document}